\documentclass[12pt]{iopart}
\expandafter\let\csname equation*\endcsname\relax
\expandafter\let\csname endequation*\endcsname\relax 

\usepackage[utf8]{inputenc}
\usepackage[T1]{fontenc}
\usepackage[ngerman, english]{babel}
\usepackage{graphicx}
\usepackage{float}

\usepackage{cite}
\usepackage{amsmath}
\usepackage{verbatim}

\renewcommand{\theequation}{\thesection.\arabic{equation}}

\newcommand{\Eqref}[1]{Eq.~(\ref{#1})}
\newcommand{\Figref}[1]{Fig.~\ref{#1}}
\newcommand{\Secref}[1]{Sec.~\ref{#1}}

\newcommand{\flMinipage}[4]{
	\hspace*{-\parindent}
	\begin{minipage}[t]{#1\textwidth}	
		#3
	\end{minipage}
	\hfill
	\begin{minipage}[t]{#2\textwidth}	
		\noindent
		#4
	\end{minipage}	
}

\usepackage{framed, color}
\definecolor{shadecolor}{gray}{.92}

\begin{document}

\newpage
\title{Comprehensive classification for Bose-Fermi mixtures}
\author{Christian Ufrecht$^1$, Matthias Meister$^1$, Albert Roura$^1$ and Wolfgang P. Schleich$^{1,2}$}
\address{$^1$ Institut für Quantenphysik and Center for Integrated Quantum Science and Technology (IQ$^\text{ST}$), Universität Ulm, Albert-Einstein-Allee
11, D-89081, Germany }
\address{$^2$ Texas A\&M University Institute for Advanced Study (TIAS), Institute for Quantum Science and Engineering (IQSE), and Department of
Physics and Astronomy, Texas A\&M University, College Station, TX 77843-4242, USA}
\ead{christian.ufrecht@uni-ulm.de}
\begin{abstract}
We present analytical studies of a boson-fermion mixture at zero temperature with spin-polarized fermions. Using the Thomas-Fermi approximation for bosons and the local-density approximation for fermions,
we find a large variety of different density shapes. In the case of continuous density, we obtain analytic conditions for each configuration for attractive as well as repulsive boson-fermion interaction. Furthermore, we analytically show that all the scenarios we describe are 
 minima of the grand-canonical 
energy functional. Finally, we provide a full classification of all possible ground states in the interpenetrative regime. Our results also apply to binary mixtures of
bosons.
\end{abstract}
%\pacs{0375}
%\submitto{\NJP}
\maketitle

\section{Introduction}
Since the experimental observation of Bose-Einstein condensation \cite{H, I}, interest 
was also oriented toward the study of ultracold fermions and mixtures of bosons and 
fermions. The theoretical  study of the density profiles of trapped boson-fermion mixtures was done, however, mainly numerically, only carried out for certain parameter 
ranges, and stability considerations lacked completeness.  In this paper we provide a comprehensive classification of all continuous ground-state profiles of a spin-polarized
 boson-fermion mixture, we will derive analytic conditions for each scenario, and rigorously show stability. 

\subsection{The richness of Bose-Fermi mixtures}
Originally, interest in ultracold fermions was, among other reasons,  spurred by the possible observation of the famous transition to the Bardeen-Cooper-Schrieffer state. Quantum degeneracy for fermionic
 systems, however, was hard to achieve as the Pauli exclusion principle forbids $s$-wave scattering in a spin-polarized sample and $p$-wave interaction is strongly suppressed at
 low temperatures \cite{P}. Therefore, a fermionic gas cannot be cooled evaporatively in a magnetic trap. This problem was circumvented by preparing the system in two spin states 
\cite{D} or by employing boson-fermion mixtures, where the Bose-Einstein condensate is cooled evaporatively and then used as a sympathetic coolant for the fermionic species \cite{A, B, J}.
 When boson-fermion mixtures with attractive interaction between the components were studied \cite{C}, a collapse of the system was observed \cite{G} whenever the number of fermions
 exceeded a certain threshold as predicted by Ref. \cite{52}. A Feshbach resonance enabled \cite{E,F} to tune the boson-fermion interaction from the collapse at attractive values via 
regimes where the densities interpenetrate to strong repulsion, where the system tends to demix. While the onset of superfluidity in a pure Fermi mixture of two spin states was proved 
in 2005 by the observation of quantized vortices \cite{Q}, it was not until 2014 that the simultaneous superfluidity of a boson-fermion mixture was observed at unitarity \cite{80}.\\ 
 A lot of effort has also been devoted to boson-fermion mixtures on the theoretical side to analyze the phase diagram. Describing the bosons by the Gross-Pitaevskii equation and the
 fermions within the local-density approximation, the collapse in attractive mixtures was studied numerically \cite{52, 68, 70, 58, 50, 67}, semi-analytically by a variational Gauss ansatz \cite{69,64},
 or in the Thomas-Fermi approximation \cite{55}. Scaling laws to determine the boundaries in parameter space between collapse, mixing, and phase separation were obtained as functions 
of the chemical potential. Their  dependence on the particle numbers, however, could only be established numerically or by rough estimates.\\ Repulsive interaction leads to a large
 variety of stable density configurations in the mixed regime, some of which were found numerically in Refs. \cite{53, 54, 62, 71}. In addition, many interesting cases in the phase-separated
 region, e.g.~mirror-symmetry breaking or the `boson burger' in an anisotropic trap were investigated in Ref. \cite{53}. The discussion was complemented by further `exotic' configurations
 found at higher energies \cite{62}. A thorough stability consideration was made for the homogeneous case \cite{63} but has not been generalized to trapped systems yet. Finally, the 
influence of finite temperature  \cite{51, 56} and beyond mean-field effects \cite{57, 66, 59, 74} have also been investigated.

\subsection{Methodology}
\begin{figure}[h] 
	\begin{center}
	\includegraphics{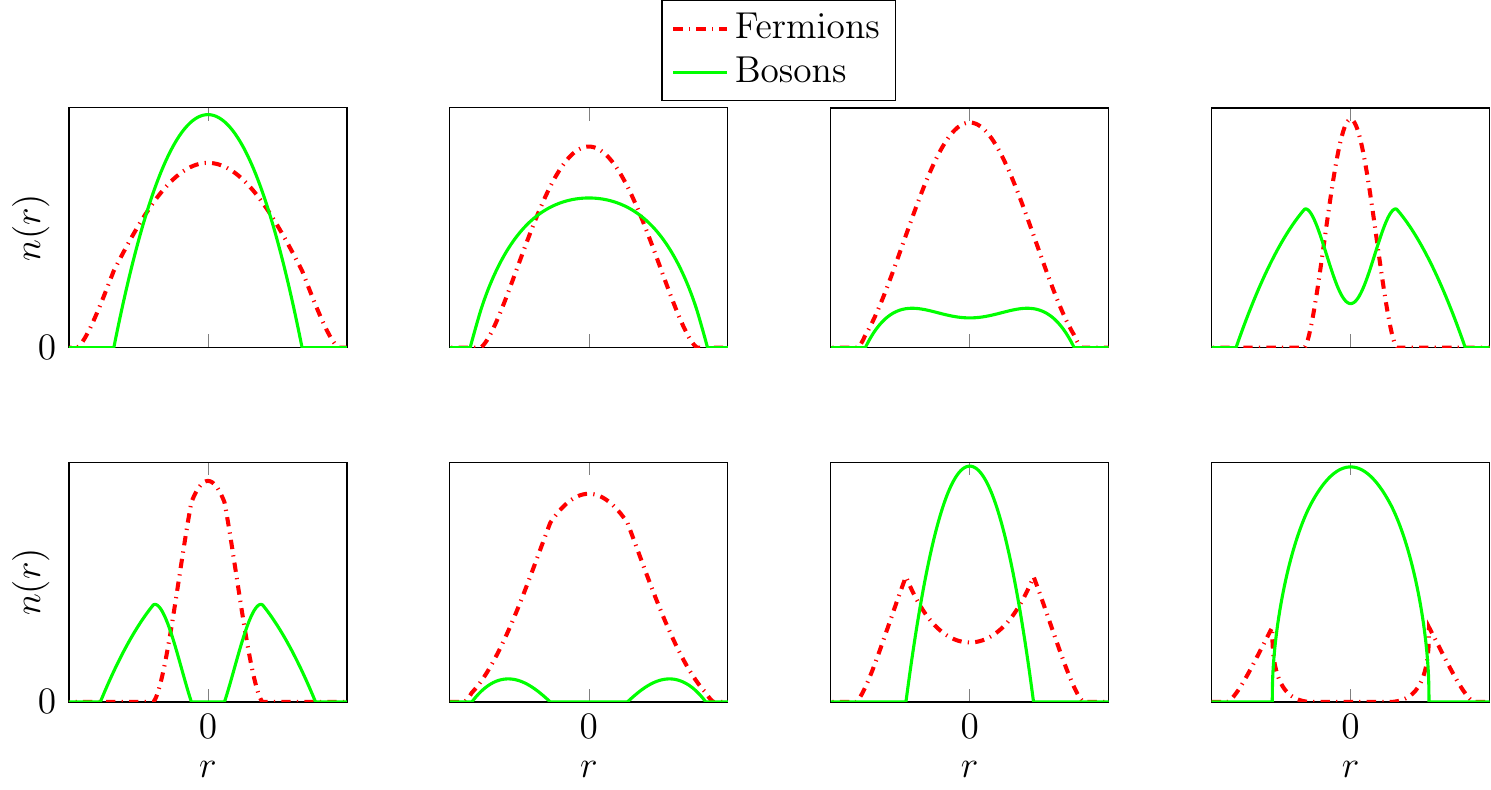}
	\caption{Possible density profiles within our classification scheme for repulsive interaction. Analytical conditions for every configuration and their detailed derivation will be presented in subsequent sections.
		The parameters involved are: The chemical potentials, the confinement for each species, and the interaction strength among the particles.}
	\label{fig:Bild1}
\end{center}
\end{figure}

\begin{figure}[h] 
	\begin{center}
                \includegraphics{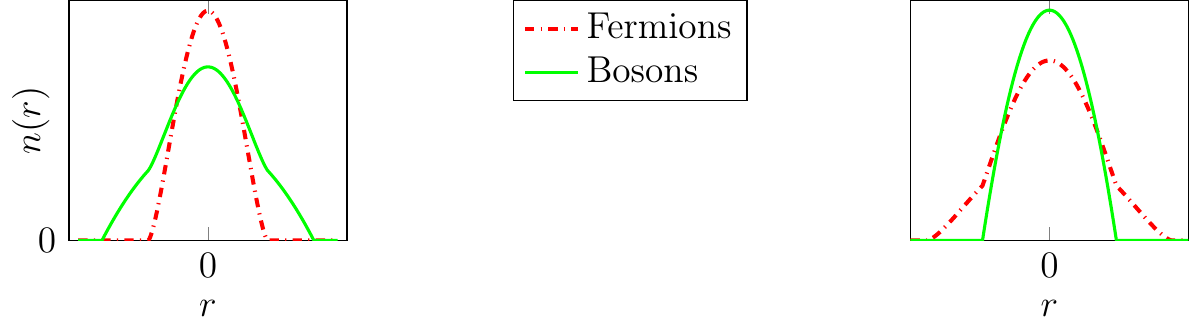}
		\caption{Possible density profiles within our classification scheme for attractive interaction.}
		\label{fig:Bild2}
	\end{center}
\end{figure}

In this paper we consider a  mixture of a Bose-Einstein condensate with fermions, which are either spin polarized or at unitarity. We provide a comprehensive classification of possible ground states in the regime where the densities are 
continuous. Configurations with discontinuities, for example when phase separation occurs, will only be discussed briefly. In our classification of density shapes we group the large 
variety of different configurations with respect to characteristic features, e.g.~the number of maxima in the densities or  which species encloses the other, similarly to what has
 been done for a binary mixture of Bose-Einstein condensates in \cite{77} and \cite{89}. A summary of our results can be found in \Figref{fig:Bild1} for repulsive interaction and 
in \Figref{fig:Bild2} for attractive interaction.

\noindent
Working in the Thomas-Fermi approximation for bosons \cite{73}
and the local-density approximation for fermions \cite{60}, the equations for the densities could in principle be solved analytically by means of Cardano's formula. It is, however, much more instructive to deduce simple but 
rigorous algebraic conditions which determine the qualitative density shapes by understanding the behavior of the equations for a given set of parameters. The discussion will be supplemented with a thorough stability consideration. Indeed, we 
will rigorously show stability for all the configurations we find. The conditions contain the chemical potentials $\mu_\mathrm{F}$ and $\mu_\mathrm{B}$ as parameters rather than the particle
 numbers $N_\mathrm{F}$ and $N_\mathrm{B}$. We will not be able to provide an analytic relation between $N$ and $\mu$ due to the nonlinearity of the equations. For that, one still has to resort to
 numerical methods. Moreover, it should be noted that conditions are stated using $>,<$ instead of $\geq,\leq$ since the subset of parameters where the real equalities hold has zero measure.\\
Finally, let us comment upon the reliability of our model. While the local-density approximation precisely describes the fermionic density even for a small number of particles, the Thomas-Fermi 
approximation is only expected to provide accurate results when the bosonic density behaves smoothly and the particle number is not too small. In the vicinity of sharp edges, which e.g.~occur in the case of phase separation or at the surface of the cloud,
the density will be smoothed out as predicted by the Gross-Pitaevskii equation \cite{78}, but the Thomas-Fermi approximation is still expected to provide qualitatively correct results.\\
 Furthermore, it should be emphasized that by a certain choice of parameters our model also reproduces all the possible density configurations for Bose-Bose mixtures.\\

\subsection{Overview}
The paper is structured as follows. In \Secref{Minima of the grand-canonical energy} we minimize the grand-cannonical energy functional and obtain nonlinear equations from which the density 
distributions follow. The structure of the equations allows several solution branches. In \Secref{Stability} we derive stability conditions to determine which of those are stable 
and obtain conditions for the number of solutions.  We then turn to the shape of the solutions. By using geometric arguments we first study the fermionic density profiles in \Secref{Classification of fermionic solutions}.
Using the results of \ref{Qualitative shape of bosonic solutions}, we also establish the bosonic profiles and provide a full classification of all possible ground-state density configurations for the boson-fermion mixture in 
\Secref{FullCalssi}.

\section{Minima of the grand-canonical energy}
\label{Minima of the grand-canonical energy}
In this section we compute the first and second functional derivative of the energy functional and obtain equations for the density shapes as well as conditions for stability.

\subsection{Energy functional}
We start our consideration with the mean-field energy functional 
\begin{align}
\nonumber
E[n_\mathrm{B}, n_\mathrm{F}]=\int\mathrm{d}^3\mathbf{r}\bigg(\frac{1}{2}g_\mathrm{B} n^2_\mathrm{B}(\mathbf{r})+V_\mathrm{B}(\mathbf{r}) n_\mathrm{B}(\mathbf{r})&+\frac{1}{\gamma+1}\kappa n_\mathrm{F}(\mathbf{r})^{\gamma+1}+V_\mathrm{F}(\mathbf{r}) n_\mathrm{F}(\mathbf{r})\\
\label{EnergyFunctionalGeneral}
&+g _\mathrm{BF}n_\mathrm{F}(\mathbf{r}) n_\mathrm{B}(\mathbf{r})\bigg)\;,
\end{align}
of a boson-fermion mixture in the Thomas-Fermi approximation for bosons and in the local-density approximation for fermions. We have introduced the real parameter $\gamma\in(0,1)$ as a generalization. If $\gamma=2/3$, we obtain the energy functional of a spin polarized boson-fermion mixture at quantum degeneracy
with $\kappa=\frac{\hbar^2}{2m_\mathrm{F}}(6\pi^2)^{2/3}$. Likewise, for a mixture of superfluid bosons and fermions at unitarity  $\kappa=\xi\frac{\hbar^2}{2m_\mathrm{F}}(3\pi^2)^{2/3}$ \cite{R},
where $\xi$ is a dimensionless renormalization factor.
Conversely, $\gamma\rightarrow 1$ describes a boson-boson mixture where $\kappa$ is the particle-particle interaction parameter of one bosonic species. 
The fermionic and bosonic potentials, $V_\mathrm{B}$ and $V_\mathrm{F}$, can be very general. The minimal requirement is that $V_\mathrm{F}=\alpha_\mathrm{F}V$ and $V_\mathrm{B}=\alpha_\mathrm{B}V$, where $\alpha_\mathrm{B},\alpha_\mathrm{F}>0$
are constant factors and $V$ is monotonic in the sense that for any given direction $\mathbf{d}$ the function
\begin{equation}
\tilde{V}(r)=V(r\mathbf{d})
\end{equation}
is monotonic for $r>0$.
Additionally, we choose $V(0)=0$ at $\mathbf{r}=0$ and $V$ to be continuous.\\
$n_\mathrm{B}$ and $n_\mathrm{F}$ are 
the bosonic and fermionic particle densities, $g_\mathrm{B}$ the bosonic particle-particle interaction parameter, and $g_\mathrm{BF}$ accounts for
the interspecies interaction.\\
To keep notation simple, we will omit the spatial coordinate $\mathbf{r}$ throughout the whole paper.\\

\subsection{First derivative of the energy functional}
We will now calculate the first functional derivative of \Eqref{EnergyFunctionalGeneral} in order to obtain equations for the density configurations.
Assuming chemical equilibrium with particle reservoirs, the state of the system minimizes the grand-canonical energy
\begin{equation}\label{DefEnergyFunc}
K[n_\mathrm{B}, n_\mathrm{F}]=\int\mathrm{d}^3\mathbf{r}\,\mathcal{K}=\int\mathrm{d}^3\mathbf{r}\,\left(\mathcal{E}-\mu_\mathrm{B}n_\mathrm{B}-\mu_\mathrm{F}n_\mathrm{F}\right)
\end{equation}
 at zero temperature, where the energy density $\mathcal{E}$ is given by the integrand of \Eqref{EnergyFunctionalGeneral}.
In order to perform the minimization, we first split the integral into three parts
$\Omega_\mathrm{B}=\left\{\mathbf{r}: n_\mathrm{B}\neq0, n_\mathrm{F}=0\right\}$, 
$\Omega_\mathrm{F}=\left\{\mathbf{r}: n_\mathrm{B}=0, n_\mathrm{F}\neq0\right\}$, and 
$\Omega_\mathrm{BF}=\left\{\mathbf{r}: n_\mathrm{B}\neq0, n_\mathrm{F}\neq0\right\}$ to characterize regions where either the bosonic or fermionic density equals zero or both species coexist. Thus, the
first variation of \Eqref{DefEnergyFunc} reads
\begin{align}
\nonumber
\delta K[n_\mathrm{B}, n_\mathrm{F}]&=\delta K_{\Omega_\mathrm{BF}}+\delta K_{\Omega_\mathrm{F}}+\delta K_{\Omega_\mathrm{B}}\\
\nonumber
&=\int_{\Omega_\mathrm{BF}}\mathrm{d}^3\mathbf{r}\left(\frac{\partial  \mathcal{K}}{\partial n_\mathrm{B}}\delta n_\mathrm{B}+\frac{\partial  \mathcal{K}}{\partial n_\mathrm{F}}\delta n_\mathrm{F}\right)\\
\nonumber
&+\int_{\Omega_\mathrm{F}}\mathrm{d}^3\mathbf{r}\left.\left(\frac{\partial  \mathcal{K}}{\partial n_\mathrm{B}}\delta n_\mathrm{B}+\frac{\partial  \mathcal{K}}{\partial n_\mathrm{F}}\delta n_\mathrm{F}\right)\right|_{n_\mathrm{B}=0}\\
&+\int_{\Omega_\mathrm{B}}\mathrm{d}^3\mathbf{r}\left.\left(\frac{\partial  \mathcal{K}}{\partial n_\mathrm{B}}\delta n_\mathrm{B}+\frac{\partial  \mathcal{K}}{\partial n_\mathrm{F}}\delta n_\mathrm{F}\right)\right|_{n_\mathrm{F}=0}\;.
\end{align}
Using \Eqref{EnergyFunctionalGeneral}, we obtain the relations
\begin{alignat}{3}
\label{deriveMinimum1}
\nonumber
&\delta K_{\Omega_\mathrm{BF}}&&=\int_{\Omega_\mathrm{BF}}&&\mathrm{d}^3\mathbf{r}\left(g_\mathrm{B}n_\mathrm{B}+V_\mathrm{B}+g_\mathrm{BF} n_\mathrm{F}-\mu_\mathrm{B}\right)\delta n_\mathrm{B}\\ 
& \quad&&+\int_{\Omega_\mathrm{BF}}&&\mathrm{d}^3\mathbf{r}\left(\kappa n_\mathrm{F}^\gamma+V_\mathrm{F}+g_\mathrm{BF} n_\mathrm{B}-\mu_\mathrm{F}\right)\delta n_\mathrm{F}\;,\\ 
\label{deriveMinimum4}
&\delta K_{\Omega_\mathrm{F}}&&=\int_{\Omega_\mathrm{F}}&&\mathrm{d}^3\mathbf{r}\left(V_\mathrm{B}+g_\mathrm{BF} n_\mathrm{F}-\mu_\mathrm{B}\right)\delta n_\mathrm{B}+\int_{\Omega_\mathrm{F}}\mathrm{d}^3\mathbf{r}\left(\kappa n_\mathrm{F}^\gamma+V_\mathrm{F}-\mu_\mathrm{F}\right)\delta n_\mathrm{F}\;,\\ 
\label{deriveMinimum5}
&\delta K_{\Omega_\mathrm{B}}&&=\int_{\Omega_\mathrm{B}}&&\mathrm{d}^3\mathbf{r}\left(g_\mathrm{B} n_\mathrm{B}+ V_\mathrm{B} -\mu_\mathrm{B}\right)\delta n_\mathrm{B}+\int_{\Omega_\mathrm{B}}\mathrm{d}^3\mathbf{r}\left(V_\mathrm{F}+g_\mathrm{BF} n_\mathrm{B} -\mu_\mathrm{F}\right)\delta n_\mathrm{F}\;.
\end{alignat}
We now search for $n_\mathrm{B}$ and $n_\mathrm{F}$ such that
\begin{equation}\label{Minimumcondition}
\delta K[n_\mathrm{B}, n_\mathrm{F}]\geq0
\end{equation}
for every infinitesimal $\delta n _\mathrm{B}$, $\delta n _\mathrm{F}$.
We note that $\delta n(\mathbf{r}_1)$ is independent of
$\delta n(\mathbf{r}_2)$ for $\mathbf{r}_1\neq\mathbf{r}_2$, so \Eqref{Minimumcondition} has to be fulfilled in every region $\Omega_\mathrm{F}$, $\Omega_\mathrm{B}$,
and $\Omega_\mathrm{BF}$ independently.\\
Let us disregard for the moment the first term in \Eqref{deriveMinimum4} and the second in \Eqref{deriveMinimum5}. We will turn to them in \Secref{Secondderivative}.
Choosing
\begin{alignat}{5}
\label{EqsForGroundstate1}
&\mu_\mathrm{B}&&=V_\mathrm{B} + g_\mathrm{B} n_\mathrm{B} + g_\mathrm{BF} n_\mathrm{F}&&\quad\quad\quad\quad\quad\quad &&\mathbf{r}\in\Omega_\mathrm{BF}\;,\\ 
\label{EqsForGroundstate2}
&\mu_\mathrm{F}&&= V_\mathrm{F} + \kappa n_\mathrm{F}^\gamma + g_\mathrm{BF} n_\mathrm{B} &&\quad\quad\quad\quad\quad\quad&&\mathbf{r}\in\Omega_\mathrm{BF}\;,\\
\label{EqsForGroundstate3}
&\mu_\mathrm{B}&&= V_\mathrm{B} + g_\mathrm{B} n_\mathrm{B}&&\quad\quad\quad\quad\quad\quad&&\mathbf{r}\in\Omega_\mathrm{B}\;,\\ 
\label{EqsForGroundstate4}
&\mu_\mathrm{F}&&= V_\mathrm{F} + \kappa n_\mathrm{F}^\gamma&&\quad\quad\quad\quad\quad\quad&&\mathbf{r}\in\Omega_\mathrm{F}\,,
\end{alignat}
then leads to
\begin{equation}
\delta K[n_\mathrm{B}, n_\mathrm{F}]=0\,,
\end{equation}
so that the solutions of these equations make \Eqref{DefEnergyFunc} stationary.

\subsection{Second derivative of the energy functional}
\label{Secondderivative}
In order to be a minimum, the second variation has to be greater than zero, which yields the expressions
\begin{alignat}{3}
\label{MaximumBedingung}
0 < &\int_{\Omega_\mathrm{F}} && &&\mathrm{d}^3\mathbf{r}\,\gamma\kappa \,n_\mathrm{F}^{\gamma-1}\,(\delta n_\mathrm{F})^2\;,\\
\label{MaximumBedingung1}
0 < &\int_{\Omega_\mathrm{B}} && &&\mathrm{d}^3\mathbf{r}\,g_\mathrm{B}\,(\delta n_\mathrm{B})^2\;,\\
\label{MaximumBedingung2}
0 < &\int_{\Omega_\mathrm{BF}} && &&\mathrm{d}^3\mathbf{r}\begin{pmatrix} \delta n_\mathrm{B} & \delta n_\mathrm{F} \end{pmatrix}\begin{pmatrix}g_\mathrm{B} & g_\mathrm{BF}\\  g_\mathrm{BF}& \gamma\kappa n_\mathrm{F}^{\gamma-1} \end{pmatrix}\begin{pmatrix} \delta n_\mathrm{B}\\ \delta n_\mathrm{F} \end{pmatrix}\;.
\end{alignat}
Note that \Eqref{MaximumBedingung} is always satisfied as $\kappa$ and $(\delta n_\mathrm{F})^2$ are always greater than zero. From \Eqref{MaximumBedingung1} we deduce that 
\begin{equation}\label{Bedingung_g}
g_\mathrm{B}>0
\end{equation} must hold true since $(\delta n_\mathrm{B})^2>0$. Therefore, stable configurations only exist for repulsive interaction among the bosons. The expression \Eqref{MaximumBedingung2} is fulfilled
for every possible infinitesimal variation of the densities exactly if the matrix is positive definite, which is true exactly if the leading principal minors are greater than zero, since the matrix is symmetric. This again provides us with $g_\mathrm{B}>0$ together with the condition
\begin{equation}\label{Bedingung_nF}
n_\mathrm{F}<\left(\frac{\gamma g_\mathrm{B}\kappa}{g_\mathrm{BF}^2}\right)^\frac{1}{1-\gamma}\;\;\;\;\;\;\;\;\;\;\;\;\;\forall\mathbf{r}\in\Omega_\mathrm{BF}\,,
\end{equation} 
which imposes a constraint on the fermionic density.\\ 
We now turn to the first contribution in \Eqref{deriveMinimum4}. 
The bosonic density is zero in $\Omega_\mathrm{F}$, therefore only $\delta n_\mathrm{B}\geq0$ for all $\mathbf{r}\in\Omega_\mathrm{F}$ are valid density variations, so that the 
bosonic density corresponds to a physical density which is larger than zero. The same applies to the fermionic density variation for the second term in \Eqref{deriveMinimum5}.
Because of that, \Eqref{Minimumcondition} is satisfied by the additional requirement
\begin{alignat}{5}
\label{AdditionalCond1}
& V_\mathrm{B}&&+g_\mathrm{BF} n_\mathrm{F}-\mu_\mathrm{B}\geq0 &&\quad\quad\quad\quad\quad\quad&&\forall\mathbf{r}\in\Omega_\mathrm{F}\\ 
\label{AdditionalCond2}
& V_\mathrm{F}&&+g_\mathrm{BF} n_\mathrm{B} -\mu_\mathrm{F}\geq0 &&\quad\quad\quad\quad\quad\quad&&\forall\mathbf{r}\in\Omega_\mathrm{B}\;.
\end{alignat}
Hence, there exist density variations where \Eqref{Minimumcondition} is true with the $>$-sign. 
In summary we have shown that the solutions of \Eqref{EqsForGroundstate1}-(\ref{EqsForGroundstate4}), (\ref{Bedingung_g}), (\ref{Bedingung_nF}), (\ref{AdditionalCond1}) and \Eqref{AdditionalCond2} 
correspond to 
a minimum of the grand-canonical energy functional, which, however, due to \Eqref{AdditionalCond1} and \Eqref{AdditionalCond2} do not necessarily make \Eqref{DefEnergyFunc} stationary.\\
It is easy to recognize that the solutions correspond to a minimum if and only if these conditions hold true because if only one is violated, one can always choose
 $\delta n_\mathrm{B}$, $\delta n_\mathrm{F}$ such that \Eqref{Minimumcondition}
is violated, too and therefore the densities do not correspond to a minimum anymore.\\
\vspace{0.15cm}\\
We conclude this section by noting that the stability considerations made above are valid for any topology of $\Omega_\mathrm{BF}$, $\Omega_\mathrm{B}$, and $\Omega_\mathrm{F}$.
In principle, one can think of three classes of solutions:\\
The first are configurations with $\Omega_\mathrm{BF}$ being the empty set, i.e.~the two species do not interpenetrate each other.
In this case we have to solve \Eqref{EqsForGroundstate3} on $\Omega_\mathrm{B}$ and \Eqref{EqsForGroundstate4} on $\Omega_\mathrm{F}$, which in general lead to discontinuities in the densities.
One can also think of a second class of solutions where $\Omega_\mathrm{BF}$ exists but there still is at least one discontinuity.  
The densities calculated for these two classes, however, will only be local minima of $K$ when (\ref{AdditionalCond1}) and \Eqref{AdditionalCond2} are satisfied, that is, only for certain parameter 
ranges and topologies of the pure regions. 
We will refer to every stable, discontinuous configuration with a coexisting region as `partially phase separated' and as `fully phase separated' if there is no such region.
In  \ref{AppendixNoPhaseSeparation} it is shown that in the case of $g_\mathrm{BF}<0$ phase separation cannot occur.
A further discussion of this interesting phenomenon, however, is not within the main
scope of this paper.\\We want to focus on the third class of solutions, that is, configurations with continuous densities. These are constructed as follows.
For a given set of parameters we first solve \Eqref{EqsForGroundstate1} and (\ref{EqsForGroundstate2}). If existent,  $\Omega_\mathrm{BF}$ is given by the region where both densities are larger than zero.
We also obtain the boundaries of this region, which is where one of the densities first reaches zero. We then use \Eqref{EqsForGroundstate3} and (\ref{EqsForGroundstate4}), respectively, to calculate 
the densities in the regions where only one particle species exists. Finally, we have 
to disregard all solutions which do not agree with  \Eqref{Bedingung_nF}. As it will turn out, this condition can be used to select the stable solution branch out of all possible solutions. We will furthermore show in \ref{Stability in pure regions} that every density configuration constructed in this manner 
is a local minimum of $K$, that is, the conditions (\ref{AdditionalCond1}) and (\ref{AdditionalCond2}) will be automatically satisfied.

\section{Existence and number of real solutions}
\label{Stability}
In the previous section we have derived conditions from the energy functional, which allow us to calculate density distributions for a given set of parameters. In this section we will examine
the nonlinear equations for the density configurations, derive a simple criterion for the stability of a solution branch, and determine conditions for the number of possible solutions.

\subsection{Selection of the stable solution branch}
\begin{figure}[H]
\begin{center}
\includegraphics{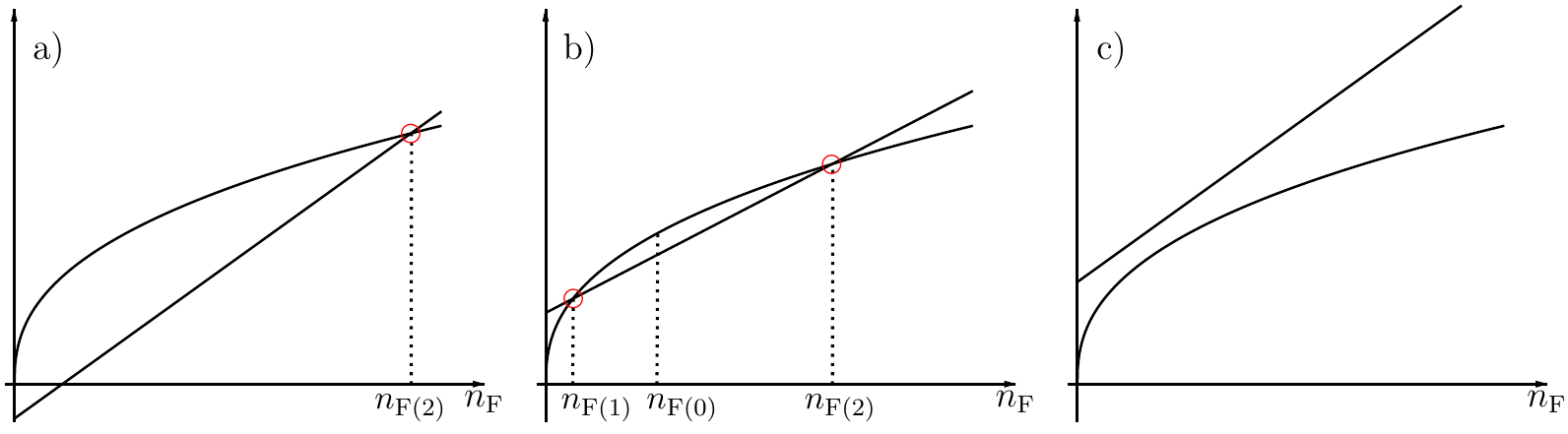}
\caption{Geometric method to solve \Eqref{DecoupledGeneral1}. Plot of the right-hand side (straight line) and the left-hand side (concave function) as a 
function of $n_\mathrm{F}$. The points where these two functions intersect correspond to the solutions of \Eqref{DecoupledGeneral1}. There are three possible scenarios depending 
on the set of parameters: One solution: Fig.~\ref{FermionicScenariosSchnittpunkte}.a, two solutions: Fig.~\ref{FermionicScenariosSchnittpunkte}.b, and no solution:
 Fig.~\ref{FermionicScenariosSchnittpunkte}.c.}
\label{FermionicScenariosSchnittpunkte}
\end{center}
\end{figure}

 In order to distinguish stable 
 solutions of \Eqref{EqsForGroundstate1} and \Eqref{EqsForGroundstate2} from those which are unphysical, we will now elucidate how to use the constraint \Eqref{Bedingung_nF} on the fermionic density.\\
By substituting  \Eqref{EqsForGroundstate1} and (\ref{EqsForGroundstate2}) into each other, the two equations can be decoupled as follows:
\begin{align}
\label{DecoupledGeneral1}
&n_\mathrm{F}^\gamma=\frac{1}{\kappa}\left( \mu_\mathrm{F}-\frac{g_\mathrm{BF}}{g_\mathrm{B}}\mu_\mathrm{B}\right)  +\frac{1}{\kappa}\left(  \frac{g_\mathrm{BF}}{g_\mathrm{B}}\alpha_\mathrm{B}-\alpha_\mathrm{F}\right)V  +\frac{g_\mathrm{BF}^2}{g_\mathrm{B}\kappa}n_\mathrm{F}\\
\label{DecoupledGeneral2}
&\frac{\kappa^\frac{1}{\gamma}}{g_\mathrm{BF}}\left(\mu_\mathrm{B}-V_\mathrm{B}\right)-\frac{\kappa^\frac{1}{\gamma}g_\mathrm{B}}{g_\mathrm{BF}}n_\mathrm{B}=\left(\mu_\mathrm{F}-V_\mathrm{F}-g_\mathrm{BF}n_\mathrm{B}\right)^\frac{1}{\gamma}.   
\end{align}
As the equations are nonlinear in the densities, we expect several solution branches for a given set of parameters. Obviously, not every solution is positive or even real, so we do not necessarily find 
physical solutions in every case. By means of geometric arguments we will demonstrate that there are always none, one, or two real solutions to both \Eqref{DecoupledGeneral1} 
and (\ref{DecoupledGeneral2}). Furthermore, we will derive criteria for deciding which corresponds to a minimum of \Eqref{EnergyFunctionalGeneral}. To do this, we consider both sides of \Eqref{DecoupledGeneral1} as functions of $n_\mathrm{F}$. Then, for a given value of $\mathbf{r}$ there exists a real solution
 when the concave function $n_\mathrm{F}^\gamma$, corresponding to the left-hand side, intersects with the straight line, the right-hand side of the equation. This can be seen 
in \Figref{FermionicScenariosSchnittpunkte}.

\noindent

\noindent
Since $\frac{g_\mathrm{BF}^2}{g_\mathrm{B}\kappa}>0$, the straight line always has positive slope and there are exactly three possible scenarios. There are either zero, one or two
 intersection points and therefore either no, one or two real solutions to \Eqref{DecoupledGeneral1}. The same conclusion can be drawn for \Eqref{DecoupledGeneral2}. Given a solution $n_\mathrm{F}$, it can be substituted 
in \Eqref{EqsForGroundstate1} yielding a real solution for $n_\mathrm{B}$ as well.\\
We now determine, which of those solutions are physically stable and which are not.  To do this, we first turn to the situation where there are two intersection points,
which we shall denote by $n_{\mathrm{F}{(1)}}$ and $n_{\mathrm{F}{(2)}}$. This situation is visualized in Fig.~\ref{FermionicScenariosSchnittpunkte}.b.  The mean 
value theorem states that there always exists a $n_{\mathrm{F}{(0)}}\in\left(n_{\mathrm{F}{(1)}},n_{\mathrm{F}{(2)}}\right)$ such that the slope of the tangent to the concave function, 
given by the first derivative $\mathrm{d}\left(n_\mathrm{F}^\gamma\right)/\mathrm{d}n_\mathrm{F}$ at $n_\mathrm{F}=n_{\mathrm{F}(0)}$, equals the slope of the straight line. This leads to the relation 
\begin{align}
\gamma\, n_{\mathrm{F}(0)}^{\gamma-1}&=\frac{g_\mathrm{BF}^2}{g_\mathrm{B}\kappa}\\
\intertext{or, equivalently,}
n_{\mathrm{F}(0)}&=\left(\frac{\gamma g_\mathrm{B}\kappa}{g_\mathrm{BF}^2}\right)^\frac{1}{1-\gamma}\;.
\end{align}
Comparison with \Eqref{Bedingung_nF} then shows that $n_{\mathrm{F}(2)}>n_{\mathrm{F}(0)}$ is unstable.
Similarly, in the case of one intersection point $n_{\mathrm{F}(2)}$ as depicted in Fig.~\ref{FermionicScenariosSchnittpunkte}.a we rewrite \Eqref{DecoupledGeneral1} as
\begin{align}
0&=c+\frac{g_\mathrm{BF}^2}{g_\mathrm{B}\kappa}n_{\mathrm{F}(2)}-n_{\mathrm{F}(2)}^\gamma\;,
\intertext{where $c$ is the ordinate intersection of the straight line and therefore $c<0$ or, equivalently,}
-c&=\frac{g_\mathrm{BF}^2}{g_\mathrm{B}\kappa}n_{\mathrm{F}(2)}-n_{\mathrm{F}(2)}^\gamma>0\;,
\end{align}
which yields after rearranging
\begin{equation}
n_{\mathrm{F}(2)}>\left(\frac{g_\mathrm{B}\kappa}{g_\mathrm{BF}^2}\right)^\frac{1}{1-\gamma}>  \left(\frac{\gamma g_\mathrm{B}\kappa}{g_\mathrm{BF}^2}\right)^\frac{1}{1-\gamma}\;.
\end{equation}
Thus, $n_{\mathrm{F}(2)}$ is an unstable solution.\\
In summary, we have shown that if there exists exactly one solution branch, it is unstable. In the case of two solutions only the fermionic branch with the smaller value is stable. If there exist two fermionic solution branches, there are two bosonic as well because of relation
\Eqref{EqsForGroundstate1}. We now establish which bosonic branch belongs to the stable fermionic one.
 For this purpose, we denote again the different solutions by $n_{\mathrm{F}(1)}$, $n_{\mathrm{B}(1)}$ and $n_{\mathrm{F}(2)}$, $n_{\mathrm{B}(2)}$. When we substitute each solution into \Eqref{EqsForGroundstate1}
 and subtract the resulting expressions from each other, we arrive at
\begin{equation}
\label{WelcheSolutionIsStable}
0=g_\mathrm{B}\left(n_\mathrm{{B}(1)}-n_\mathrm{{B}(2)}\right)+g_\mathrm{BF}\left(n_\mathrm{{F}(1)}-n_\mathrm{{F}(2)}\right)\;.
\end{equation}
Let $n_\mathrm{{F}(1)}$ be the fermionic solution with the smaller value. For $g_\mathrm{BF}>0$ the second contribution in \Eqref{WelcheSolutionIsStable} is then 
smaller than zero. Hence, the first term is larger than zero and therefore $n_\mathrm{{B}(1)}>n_\mathrm{{B}(2)}$. Thus, the fermionic solution with the smaller value corresponds to the bosonic solution with the greater value and these solutions are stable. In contrast,
for $g_\mathrm{BF}<0$ the smaller fermionic solution corresponds to the smaller bosonic one, which are the stable solutions in this case.\\

\subsection{Conditions for the number of solutions}\label{Conditions for the number of solutions}
\begin{figure}[H]
\begin{center}
\includegraphics{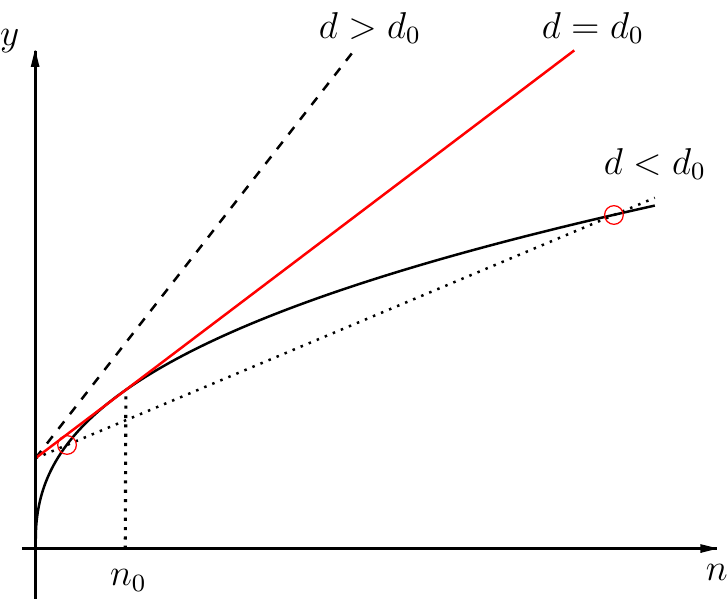}
\caption{Visualization of \Eqref{xEquationForSolutionCondition}. Plot of the left-hand side (concave function) and the right-hand side (straight line), which intersect at points that correspond to a solution.
For every $c>0$ there exists a single straight line (solid line) with slope $d_0$ which is 
tangent to the concave function. If $d<d_0$, there are two solutions (dotted line) and if $d>d_0$, there is no solution (dashed line).}
\label{DerivationSolCondition}
\end{center}
\end{figure}

The next step is to derive conditions which determine the number of real solutions to \Eqref{DecoupledGeneral1} and (\ref{DecoupledGeneral2}). We start our analysis by introducing the following abbreviations
in \Eqref{DecoupledGeneral1}:
\begin{equation}
\label{Identification_a_and_c}
d=\frac{g_\mathrm{BF}^2}{g_\mathrm{B}\kappa}\;\;\;\;\text{and}\;\;\;\;c=\frac{1}{\kappa}\left( \mu_\mathrm{F}-\frac{g_\mathrm{BF}}{g_\mathrm{B}}\mu_\mathrm{B}\right)+\frac{1}{\kappa}\left(\frac{g_\mathrm{BF}}{g_\mathrm{B}}\alpha_\mathrm{B}-\alpha_\mathrm{F}\right)V\;,
\end{equation}
which lead to
\begin{equation}
\label{xEquationForSolutionCondition}
n_\mathrm{F}^\gamma=d\,n_\mathrm{F}+c\;,
\end{equation}
and where it is important to note that $\gamma\in(0,1)$ and $d>0$. \\
As discussed above and seen from Fig.~\ref{FermionicScenariosSchnittpunkte}.a, for $c<0$ there is one positive solution.\\
Let us analyze what happens for $c>0$. We start our consideration by noting that the derivative of $n^\gamma$ with respect to $n$ is greater than zero for $n\in(0,\infty)$. Then it follows that for every $c>0$ 
there exists one $d_0>0$ such that the straight line $y=d_0\,n_\mathrm{F}+c$ is tangent to $y=n_\mathrm{F}^\gamma$ at $n_\mathrm{F}=n_{\mathrm{F}(0)}$ as can be seen in \Figref{DerivationSolCondition}. Thus, we find
\begin{align}
\label{ZeroSolutions}
&      \text{no real solution for } && d>d_0\\
\label{TwoSolutions}
&    \text{and two positive solutions for } && d<d_0\;. 
\end{align}
When we solve the system of equations
\begin{align}
\label{onepositive}
\gamma n_{\mathrm{F}(0)}^{\gamma-1}&=d_0\;,\\
n_{\mathrm{F}(0)}^\gamma&=d_0n_{\mathrm{F}(0)}+c\;,
\end{align}  
we obtain the explicit expression
\begin{equation}
\label{BedingungSolutions1}
d_0=\gamma\left(\frac{1-\gamma}{c}\right)^\frac{1-\gamma}{\gamma}\;.
\end{equation}  
Finally, the combination of \Eqref{BedingungSolutions1} with \Eqref{ZeroSolutions} and (\ref{TwoSolutions}) leads after rearranging terms to the conditions
\begin{alignat}{5}
\label{ConditionsForSolutions0}
\left(\frac{\gamma}{d}\right)^\frac{\gamma}{1-\gamma}\left(1-\gamma\right)\;&& &&<\;& c  \quad\quad&&\text{for no real solution,}\\
\label{ConditionsForSolutions2}
\left(\frac{\gamma}{d}\right)^\frac{\gamma}{1-\gamma}\left(1-\gamma\right)\;&& &&>\;& c>0 \quad\quad&& \text{for two positive solutions, and}\\
\label{ConditionsForSolutions1}
&& && &c< 0 \quad\quad&&\text{for one positive solution.}
\end{alignat}
\noindent
Substituting back the expressions for $d$ and $c$ according to \Eqref{Identification_a_and_c}, the three cases become
\begin{alignat}{3}
\intertext{- no real solution if}
\label{Solutionsattheorigin1}
&\chi\left(1-\gamma\right)<&&\left(\mu_\mathrm{F}-\frac{g_\mathrm{BF}}{g_\mathrm{B}}\mu_\mathrm{B}\right)+\left(\frac{g_\mathrm{BF}}{g_\mathrm{B}}\alpha_\mathrm{B}-\alpha_\mathrm{F}\right)V\;,\\
\intertext{- two solutions if}
\label{Solutionsattheorigin2}
&\chi\left(1-\gamma\right)>&&\left(\mu_\mathrm{F}-\frac{g_\mathrm{BF}}{g_\mathrm{B}}\mu_\mathrm{B}\right)+\left(\frac{g_\mathrm{BF}}{g_\mathrm{B}}\alpha_\mathrm{B}-\alpha_\mathrm{F}\right)V>0\;,\\
\intertext{- one solution if}
\label{Solutionsattheorigin3}
&\quad &&\left( \mu_\mathrm{F}-\frac{g_\mathrm{BF}}{g_\mathrm{B}}\mu_\mathrm{B}\right)+\left(\frac{g_\mathrm{BF}}{g_\mathrm{B}}\alpha_\mathrm{B}-\alpha_\mathrm{F}\right)V<0\;.
\end{alignat}
Here, we have defined
\begin{equation}
\label{Definechi}
\chi:=\kappa \left(\frac{g_\mathrm{B}\kappa\gamma}{g^2_\mathrm{BF}}\right)^\frac{\gamma}{1-\gamma}
\end{equation}
for convenience.
These conditions also determine the possible solutions for the bosonic density, which follow from \Eqref{DecoupledGeneral2}. This is because given any solution for the fermionic density, the bosonic density can be easily obtained from the linear relation \Eqref{EqsForGroundstate1}. For example,
if there are two fermionic branches, there will be two bosonic ones as well.
The value of the bosonic solutions can, however, be negative, which does not correspond to a physical density. Therefore, their positivity needs to be explicitly checked. This will be done in \ref{Qualitative shape of bosonic solutions}.

\newpage
\section{Qualitative shape of the fermionic solutions}\label{Classification of fermionic solutions}
\begin{figure}[H]
\includegraphics{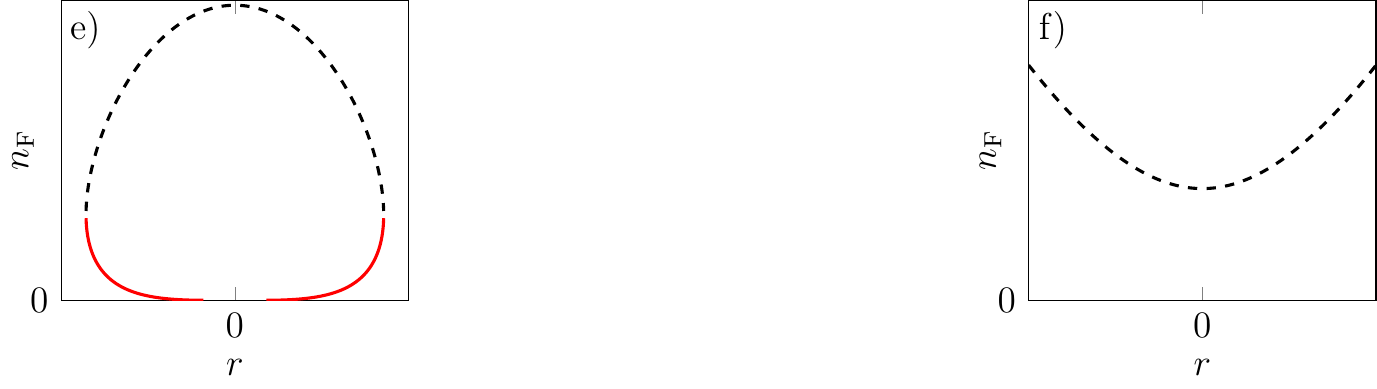}
\vspace{-17pt}
\begin{flalign*}
	&\hspace{27pt}\frac{g_\mathrm{BF}}{g_\mathrm{B}}\alpha_\mathrm{B}-\alpha_\mathrm{F}>0 \hspace{196pt} \frac{g_\mathrm{BF}}{g_\mathrm{B}}\alpha_\mathrm{B}-\alpha_\mathrm{F}<0&
\end{flalign*}
\caption{Classification of the solutions of \Eqref{DecoupledGeneral1}. The conditions \Eqref{Solutionsattheorigin1}-(\ref{Solutionsattheorigin3}) for the number of solutions 
at the origin are used in combination with geometric arguments based on the conditions \Eqref{geometricargument1} and (\ref{geometricargument2}) in order to determine the six different qualitative 
fermionic density profiles.}
\label{FermionicScenarios}
\end{figure}

At first sight these conditions may not seem very helpful. However, we will show that one can already make a full classification of the qualitative shape of the solutions to \Eqref{DecoupledGeneral1}
from the conditions that we have derived in the previous sections. To this end, we will again make use of geometric arguments. Recall from
 \Eqref{Identification_a_and_c} that $c$ depends on $V(\mathbf{r})$ and therefore on $\mathbf{r}$. Since $V$ is monotonic, an increase of $|\mathbf{r}|$ corresponds to a 
vertical shift of the straight line in \Figref{FermionicScenariosSchnittpunkte}. The line moves
\begin{align}
\label{geometricargument1}
&\text{upwards if}&&\frac{g_\mathrm{BF}}{g_\mathrm{B}}\alpha_\mathrm{B}-\alpha_\mathrm{F}>0\;\;\text{ and}\\
\label{geometricargument2}
&\text{downwards if}&&\frac{g_\mathrm{BF}}{g_\mathrm{B}}\alpha_\mathrm{B}-\alpha_\mathrm{F}<0\;.
\end{align}
At the origin $V(\mathbf{r})=0$ by definition and therefore
\begin{equation}
c=\frac{1}{\kappa}\left( \mu_\mathrm{F}-\frac{g_\mathrm{BF}}{g_\mathrm{B}}\mu_\mathrm{B}\right)\;.
\end{equation}
In \Figref{FermionicScenarios} we summarize the possible solutions and display them together with their corresponding conditions. We indicate stable density 
configurations by solid lines, while we use dashed lines for unstable solutions.\\
We illustrate how to proceed in general using the case depicted in Fig.~\ref{FermionicScenarios}.a as an example. The condition above Fig.~\ref{FermionicScenarios}.a states that there are 
two solutions at the origin. This situation is depicted by the dotted line in \Figref{DerivationSolCondition}. The condition below Fig.~\ref{FermionicScenarios}.a corresponds to an 
upwards moving straight line, that is, the value of the greater solution 
becomes smaller, while the value of the smaller solution becomes larger as we increase $|\mathbf{r}|$ and therefor $V$ until they reach the same value at the point where the straight line is tangent to the concave function. We will refer to such a point
as a `tangent point'. Here, the branches end, hence there is no solution for larger values of $|\mathbf{r}|$ anymore.\\
There is a second way for a solution branch to disappear, which can be seen for instance in Fig.~\ref{FermionicScenarios}.b. At the origin we observe two solutions again, the straight line in \Figref{DerivationSolCondition} moves downwards, the solution
with the smaller value becomes smaller until it reaches zero and then disappears. We will refer to such a point as a `ceasing point'.\\
In this spirit, we obtain all the possible different density profiles for the fermions. Recall from \Secref{Stability} that if there are two solutions the lower is stable, and if there 
is only one solution, it is unstable.

\newpage
\vspace{0.7cm}
\begin{flalign*}
\text{Stable solution at the origin:}\quad\quad\quad\quad\chi\left(1-\gamma\right)>\left( \mu_\mathrm{F}-\frac{g_\mathrm{BF}}{g_\mathrm{B}}\mu_\mathrm{B}\right)>0&&
\end{flalign*}
\vspace{-15pt}
\begin{figure}[H]
\includegraphics{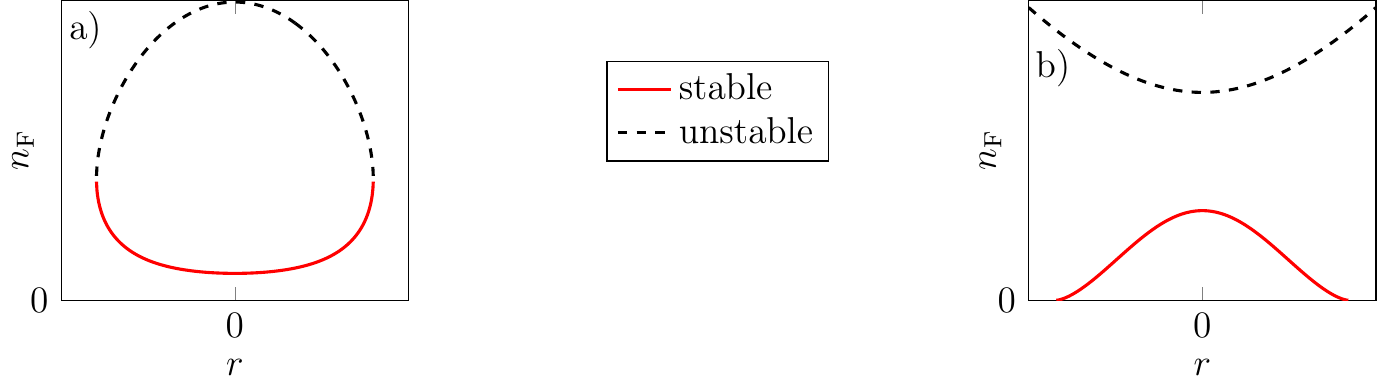}
\end{figure}
\noindent
\vspace{-40pt}
\begin{flalign*}
	&\hspace{27pt}\frac{g_\mathrm{BF}}{g_\mathrm{B}}\alpha_\mathrm{B}-\alpha_\mathrm{F}>0 \hspace{196pt} \frac{g_\mathrm{BF}}{g_\mathrm{B}}\alpha_\mathrm{B}-\alpha_\mathrm{F}<0&
\end{flalign*}

\begin{flalign*}
\text{No solution at the origin:}\quad\quad\quad\quad\quad\;\;\chi\left(1-\gamma\right)<\left( \mu_\mathrm{F}-\frac{g_\mathrm{BF}}{g_\mathrm{B}}\mu_\mathrm{B}\right) &&
\end{flalign*}
\vspace{-15pt}
\begin{figure}[H]
\includegraphics{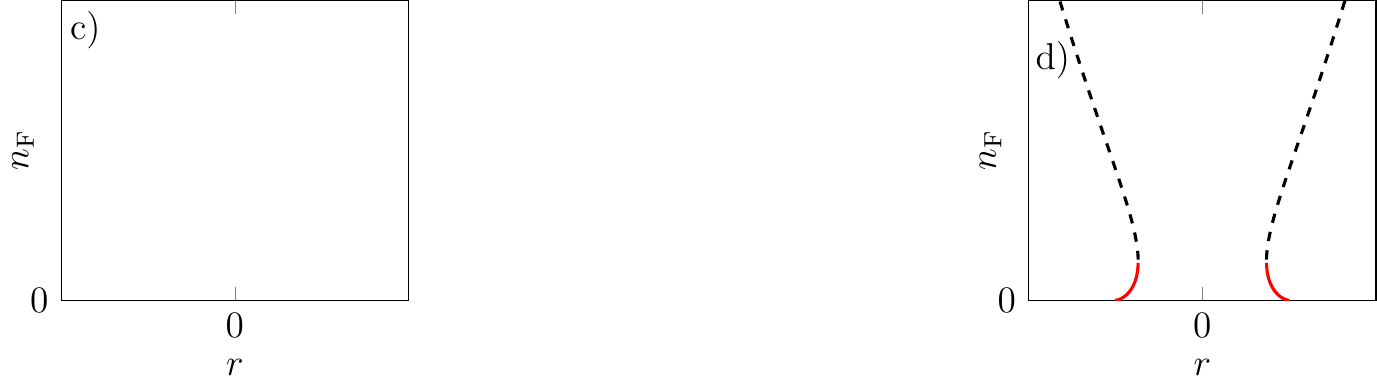}
\end{figure}
\noindent
\vspace{-40pt}
\begin{flalign*}
&\hspace{27pt}\frac{g_\mathrm{BF}}{g_\mathrm{B}}\alpha_\mathrm{B}-\alpha_\mathrm{F}>0 \hspace{196pt} \frac{g_\mathrm{BF}}{g_\mathrm{B}}\alpha_\mathrm{B}-\alpha_\mathrm{F}<0&
\end{flalign*}

\vspace{0.7cm}
\begin{flalign*}
\text{Unstable solution at the origin:}\quad\quad\quad\,\mu_\mathrm{F}-\frac{g_\mathrm{BF}}{g_\mathrm{B}}\mu_\mathrm{B}<0 &&
\end{flalign*}
\vspace{-15pt}
\noindent

\newpage
\section{Classification of density configurations}\label{FullCalssi}
In the preceding section we were able to provide a full classification of solutions to \Eqref{DecoupledGeneral1} with the help of geometric arguments.
When we apply the same reasoning to \Eqref{DecoupledGeneral2}, a problem arises, which will be discussed next.
The left-hand side of \Eqref{DecoupledGeneral2} corresponds to a straight line with negative slope for $g_\mathrm{BF}>0$ and positive slope for $g_\mathrm{BF}<0$, while the right-hand side is a convex 
function. The solutions of \Eqref{DecoupledGeneral2} are then identified by the values of $n_\mathrm{B}$ where the two functions intersect.
The next step in \Secref{Classification of fermionic solutions} was to increase the potential in \Eqref{DecoupledGeneral1} and to track the change of the intersection points in order to obtain
the qualitative, spatial dependency of the solutions. However, in \Eqref{DecoupledGeneral2} both sides depend on $V$, that is, both the convex function and the straight line change as the potential
 varies, but in possibly different directions and we cannot infer how the value
of the intersection point changes. The classification of fermionic solutions has, nevertheless, already provided us with a lot of information about the bosonic solutions. Indeed, it follows from
 \Eqref{EqsForGroundstate1} that they will exhibit ceasing and tangent points at the same locations as the fermionic solutions, just as the number of fermionic and bosonic solution branches at every point $\mathbf{r}$ is the same. 
Since the value of the bosonic solutions at these points may be negative, as can be seen from \Eqref{EqsForGroundstate1}, in order to correspond to a physical solution, its positivity has to be checked explicitly.
We will address these problems in \ref{Qualitative shape of bosonic solutions}. The calculations  in this appendix are somewhat technical, the reader who is mainly interested in the classification
of density profiles can skip the details. First, in \ref{Classification of bosonic solutions at the origin} we recall the condition for two solutions at the origin and subsequently establish further conditions to specify the sign of $n_\mathrm{B}$.
In the second step in \ref{Classification of bosonic solutions Vg0} we make analytical statements about the value of the bosonic solution at ceasing, 
tangent and stationary points (the derivative of the bosonic solution becomes zero).
These two steps will provide us with all the information we need for a full classification.\\
Before we start to discuss the different scenarios, let us first make a few general remarks. As mentioned before, 
we focus on a classification of continuous density profiles.  If the fermionic solution branch disappears at a connection point e.g.~in  Fig.~\ref{FermionicScenarios}.a, d or e, the bosonic solution has to be smaller than zero at this point. 
If this
were not the case,  it would not be possible to continuously connect the solution to the pure region, where only one species is present. Contrary to that, at a fermionic ceasing point, e.g.~in
 Fig.~\ref{FermionicScenarios}.b, d or e
the bosonic solution can be positive, if this point is the border to the pure bosonic region, or negative, then the point at which the bosonic solution branch crosses zero is the border to the
 pure fermionic solution. In the conditions we use the abbreviations
\begin{equation}
\beta=\kappa\left(\gamma\frac{\alpha_\mathrm{B}}{\alpha_\mathrm{F}}\frac{\kappa}{g_\mathrm{BF}}\right)^\frac{\gamma}{1-\gamma}\quad\quad\text{and}\quad\quad\chi=\kappa \left(\frac{g_\mathrm{B}\kappa\gamma}{g^2_\mathrm{BF}}\right)^\frac{\gamma}{1-\gamma}\;,
\end{equation}
defined in \Eqref{Definechi} and \ref{DefBeta} for better readability.

\newpage
\subsection{Classification for repulsive interaction $g_\mathrm{BF}>0$}
\label{Classification for repulsive interaction}
Keeping these remarks in mind, we proceed as described in the following. Choosing a fermionic scenario in \Figref{FermionicScenarios} defines the number of fermionic and therefore bosonic
solutions at the origin. \ref{Qualitative shape of bosonic solutions} provides further conditions, which determine their sign  at the origin and their values at ceasing, tangent,
and stationary points.

\subsubsection{Fermionic scenario Fig.~\ref{FermionicScenarios}.a and Fig.~\ref{FermionicScenarios}.b }
\hspace{1cm}\\
We start with the fermionic scenarios depicted in Fig.~\ref{FermionicScenarios}.a and \ref{FermionicScenarios}.b, which both feature two solutions at the origin. 
If the greater and therefore stable bosonic solution is larger than zero, that is, the additional conditions below Fig.~\ref{BosonenScenario1}.b or \ref{BosonenScenario1}.c are satisfied, the configuration exhibits 
a mixed region at the origin. Let us discuss \Figref{FermionicScenarios}.b first. Here, we observe a ceasing point, which will also be found in the bosonic solution. If the value of the bosonic ceasing point is smaller than zero, that is $\mu_\mathrm{F}-\frac{\alpha_\mathrm{F}}{\alpha_\mathrm{B}}\mu_\mathrm{B}>0$ in \Eqref{conditionForCeasing2}, 
the bosonic density reaches zero before the fermionic does.  The point where the bosonic density becomes zero determines the border 
of the mixed region to the pure fermionic region.
If a stationary point exists due to \Eqref{CondStationary4}, it will be found in the bosonic solution branch which corresponds to the fermionic branch with negative slope because of \Eqref{StationaryPointSlopeCondition}. This is the stable solution.
Furthermore, it follows from \Eqref{SlopeCeasing} that the slope at the ceasing point is negative in the direction of increasing $V$. As there is only
one stationary point additionally to the one at the origin, it has to be a maximum. This either leads to scenario 1.1.1 or scenario 1.1.2.
If the ceasing point in contrast is larger than zero, $\mu_\mathrm{F}-\frac{\alpha_\mathrm{F}}{\alpha_\mathrm{B}}\mu_\mathrm{B}<0$ is true. Following along the same lines as before, we can easily convince
ourselves that we find scenario 1.2.1 and scenario 1.2.2.\\
Before we discuss Fig.~\ref{FermionicScenarios}.b in combination with condition Fig.~\ref{BosonenScenario1}.a, let us turn to Fig.~\ref{FermionicScenarios}.a, where we observe a tangent point but no ceasing point, first again in combination with Fig.~\ref{BosonenScenario1}.b or \ref{BosonenScenario1}.c.
In Fig.~\ref{FermionicScenarios}.a the fermionic density does not reach zero. Consequently, in order to represent a physically reasonable configuration the bosonic tangent point has to be smaller than zero 
to delimit the mixed region. This is ensured by using \Eqref{conditionConnection3}. Since $\frac{\partial n_\mathrm{F}}{\partial V}>0$ everywhere, \Eqref{StationaryPointSlopeCondition} cannot 
be satisfied and there is no stationary point. In summary, we find scenario 1.3.\\
Let us now discuss Fig.~\ref{FermionicScenarios}.a in the context of the conditions of Fig.~\ref{BosonenScenario1}.a. It is easy to convince ourselves that there is no physical scenario with 
this fermionic density. As the stable fermionic solution does never become zero, for a physically reasonable scenario, the bosonic solution, which is smaller than zero at the origin, should increase, cross zero and then decrease
until it reaches zero again. This, however, would imply the existence of a stationary point, which  contradicts \Eqref{StationaryPointSlopeCondition}.\\
Contrary to that,  Fig.~\ref{FermionicScenarios}.b in combination with the conditions of Fig.~\ref{BosonenScenario1}.a leads to further scenarios.  At the origin the bosonic solution is smaller than zero, however, as the potential increases
it might become larger and eventually cross zero, leading to a mixed region.

\newpage
\footnotesize
\hspace{1cm}\\
\vspace{-1.6cm}
\begin{shaded}
\noindent
1. $g_\mathrm{BF}>0$, mixed region at the origin
\begin{equation*}
\chi\left(1-\gamma\right)>\left(\mu_\mathrm{F}-\frac{g_\mathrm{BF}}{g_\mathrm{B}}\mu_\mathrm{B}\right)>0 
\end{equation*}
\begin{equation*}
\frac{\kappa^\frac{1}{\gamma}\mu_\mathrm{B}}{g_\mathrm{BF}}>\mu_\mathrm{F}^\frac{1}{\gamma}\quad \text{or}\quad \left[\frac{\kappa^\frac{1}{\gamma}\mu_\mathrm{B}}{g_\mathrm{BF}}<\mu_\mathrm{F}^\frac{1}{\gamma}\quad\text{and}\quad \frac{1}{\gamma}\mu_\mathrm{F}^{\frac{1}{\gamma}-1}>\frac{\kappa^\frac{1}{\gamma}g_\mathrm{B}}{g_\mathrm{BF}^2}\right] 
\end{equation*}

\vspace{0.3cm}
\noindent
1.1 Fermionic density decreases, the bosonic density reaches zero first
\begin{equation*}
\frac{g_\mathrm{BF}}{g_\mathrm{B}}\alpha_\mathrm{B}-\alpha_\mathrm{F}<0,\quad\mu_\mathrm{F}-\frac{\alpha_\mathrm{F}}{\alpha_\mathrm{B}}\mu_\mathrm{B}>0
\end{equation*}

\vspace{0.3cm}
\hspace{-4pt}
\flMinipage{0.59}{0.41}{
\hspace{2cm}
1.1.1 No stationary point	
\vspace{-0.2cm}
\begin{flalign*}
\hspace{2.15cm}
&g_\mathrm{B}\mu_\mathrm{F}-g_\mathrm{BF}\mu_\mathrm{B}+\beta\left(g_\mathrm{BF}\gamma\frac{\alpha_\mathrm{B}}{\alpha_\mathrm{F}}-g_\mathrm{B}\right)<0&
\end{flalign*}
}{
\hspace{-8.5pt}
1.1.2 Stationary point 
\vspace{-0.2cm}
\begin{flalign*}
&g_\mathrm{B}\mu_\mathrm{F}-g_\mathrm{BF}\mu_\mathrm{B}+\beta\left(g_\mathrm{BF}\gamma\frac{\alpha_\mathrm{B}}{\alpha_\mathrm{F}}-g_\mathrm{B}\right)>0&
\end{flalign*}
}

\vspace{-0.45cm}
\begin{figure}[H]
\hspace{1.92cm}
\includegraphics{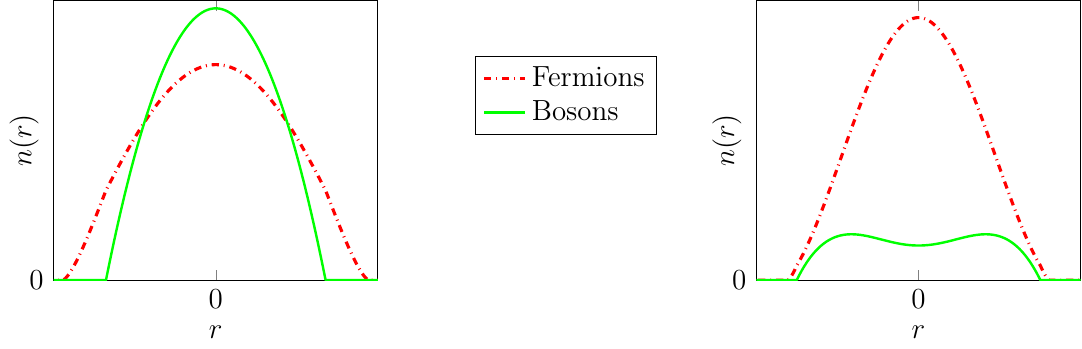}
\end{figure}
%%%%%%%%%%%%%%%%%%%%%%%%%%%%%%%%%%%%%%%%%%%%%%%%%%%%%%%%%%%%%%%%%%%%%%%%%%%%%%%%%%%%%%%%%%%%%%%%%%%%%%%%%%%%%%%%%%%%%%%%%%%
\vspace{-0.3cm}

\noindent
1.2 Fermionic density decreases and reaches zero first
\begin{equation*}
\frac{g_\mathrm{BF}}{g_\mathrm{B}}\alpha_\mathrm{B}-\alpha_\mathrm{F}<0,\quad\mu_\mathrm{F}-\frac{\alpha_\mathrm{F}}{\alpha_\mathrm{B}}\mu_\mathrm{B}<0 
\end{equation*}

\vspace{0.3cm}
\hspace{-4pt}	
\flMinipage{0.59}{0.41}{
\hspace{2cm}
1.2.1 No stationary point   
        \vspace{-0.2cm}
	\begin{flalign*}
        \hspace{2.1cm}
		&\,g_\mathrm{B}\mu_\mathrm{F}-g_\mathrm{BF}\mu_\mathrm{B}+\beta\left(g_\mathrm{BF}\gamma\frac{\alpha_\mathrm{B}}{\alpha_\mathrm{F}}-g_\mathrm{B}\right)<0&
	\end{flalign*}
	}{
\hspace{-8.5pt}
1.2.2 Stationary point
                \vspace{-0.2cm}
		\begin{flalign*}
		&g_\mathrm{B}\mu_\mathrm{F}-g_\mathrm{BF}\mu_\mathrm{B}+\beta\left(g_\mathrm{BF}\gamma\frac{\alpha_\mathrm{B}}{\alpha_\mathrm{F}}-g_\mathrm{B}\right)>0 &
		\end{flalign*}
		}
	\vspace{-0.3cm}
	\begin{figure}[H]
		\hspace{2.27cm}
\includegraphics{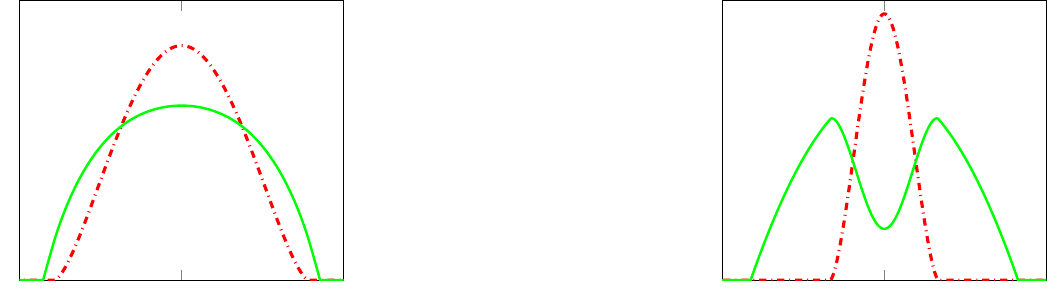}
	\end{figure}
%%%%%%%%%%%%%%%%%%%%%%%%%%%%%%%%%%%%%%%%%%%%%%%%%%%%%%%%%%%%%%%%%%%%%%%%%%%%%%%%%%%%%%%%%%%%%%%%%%%%%%%%%%%%%%%%%%%%%%%%%%%
\vspace{-0.3cm}

\noindent
1.3 Fermionic density increases in the mixed region
\begin{equation*}
\frac{g_\mathrm{BF}}{g_\mathrm{B}}\alpha_\mathrm{B}-\alpha_\mathrm{F}>0,     \quad \chi\left(\gamma\frac{g_\mathrm{B}}{g_\mathrm{BF}}-\frac{\alpha_\mathrm{B}}{\alpha_\mathrm{F}}\right)+\frac{\alpha_\mathrm{B}}{\alpha_\mathrm{F}}\mu_\mathrm{F}-\mu_\mathrm{B}<0 
\end{equation*}

\vspace{-0.3cm}
\begin{figure}[H]
\begin{center}
\hspace{-0.29cm}
\includegraphics{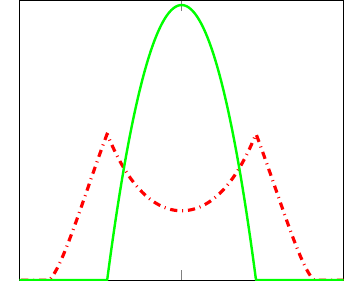}
\end{center}
\end{figure}
\vspace{-0.5cm}
\end{shaded}
%%%%%%%%%%%%%%%%%%%%%%%%%%%%%%%%%%%%%%%%%%%%%%%%%%%%%%%%%%%%%%%%%%%%%%%%%%%%%%%%%%%%%%%%%%%%%%%%%%%%%%%%%%%%%%%%%%%%%%%%%%%

\normalsize

\noindent
There are two ways to achieve that. First, we choose $\mu_\mathrm{F}-\frac{\alpha_\mathrm{F}}{\alpha_\mathrm{B}}\mu_\mathrm{B}<0$
in \Eqref{conditionForCeasing2}, so that the value of the ceasing point is larger than zero. As a result of  \Eqref{SlopeCeasing} the slope at the ceasing point is negative, leading to a stationary point. Consequently, we find scenario 2.1. Secondly, if we choose
$\mu_\mathrm{F}-\frac{\alpha_\mathrm{F}}{\alpha_\mathrm{B}}\mu_\mathrm{B}>0$ the ceasing point is below zero. In order to be physical, there has to be a stationary point, that is, Eq.~(\ref{CondStationary4}) has to be true, and its  value has to be 
above zero, which is guaranteed by \Eqref{CondStationary3}. In summary, we find scenario 2.2.

\footnotesize
\begin{shaded}
\noindent
2. $g_\mathrm{BF}>0$, zero bosonic density at the origin
\begin{equation*}
	\chi\left(1-\gamma\right)>\left(\mu_\mathrm{F}-\frac{g_\mathrm{BF}}{g_\mathrm{B}}\mu_\mathrm{B}\right)>0 
\end{equation*}
\begin{equation*}
	\frac{\kappa^\frac{1}{\gamma}\mu_\mathrm{B}}{g_\mathrm{BF}}<\mu_\mathrm{F}^\frac{1}{\gamma},\quad\frac{1}{\gamma}\mu_\mathrm{F}^{\frac{1}{\gamma}-1}<\frac{\kappa^\frac{1}{\gamma}g_\mathrm{B}}{g_\mathrm{BF}^2}
\end{equation*}
\noindent

\vspace{0.3cm}
\flMinipage{0.59}{0.41}{
\hspace{2cm}
2.1 Fermionic density reaches zero first
\vspace{-0.2cm}
\begin{flalign*}
\hspace{2.13cm}
&\frac{g_\mathrm{BF}}{g_\mathrm{B}}\alpha_\mathrm{B}-\alpha_\mathrm{F}<0,\quad\mu_\mathrm{F}-\frac{\alpha_\mathrm{F}}{\alpha_\mathrm{B}}\mu_\mathrm{B}<0 &
\end{flalign*}
}{
\hspace{-8.5pt}
2.2 Bosonic density reaches zero first
\vspace{-0.2cm}
\begin{flalign*}
&\hspace{-0.03cm} \frac{g_\mathrm{BF}}{g_\mathrm{B}}\alpha_\mathrm{B}-\alpha_\mathrm{F}<0,\quad\mu_\mathrm{F}-\frac{\alpha_\mathrm{F}}{\alpha_\mathrm{B}}\mu_\mathrm{B}>0 &\\
& \mu_\mathrm{F}-\frac{\alpha_\mathrm{F}}{\alpha_\mathrm{B}}\mu_\mathrm{B} <\beta(1-\gamma)&\\
& g_\mathrm{B}\mu_\mathrm{F}-g_\mathrm{BF}\mu_\mathrm{B}+\beta\left(g_\mathrm{BF}\gamma\frac{\alpha_\mathrm{B}}{\alpha_\mathrm{F}}-g_\mathrm{B}\right)>0 &\\
\end{flalign*}
}

\vspace{-1cm}
\begin{figure}[H]
	\hspace{1.92cm}
\includegraphics{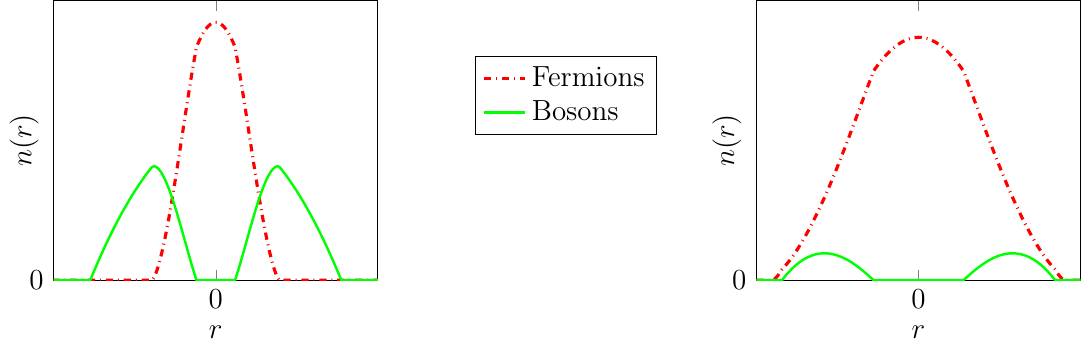}
\end{figure}
\vspace{-0.6cm}

\hspace{1.23cm}
There is a second set of parameters for scenario 2.1 and 2.2, which are

\vspace{-0.6cm}
\flMinipage{0.59}{0.41}{
\hspace{2cm}
\begin{flalign*}
\hspace{2.13cm}
&\chi\left(1-\gamma\right)<\left(\mu_\mathrm{F}-\frac{g_\mathrm{BF}}{g_\mathrm{B}}\mu_\mathrm{B}\right) &\\
&\frac{g_\mathrm{BF}}{g_\mathrm{B}}\alpha_\mathrm{B}-\alpha_\mathrm{F}<0,\quad\mu_\mathrm{F}-\frac{\alpha_\mathrm{F}}{\alpha_\mathrm{B}}\mu_\mathrm{B}<0 &\\
&\chi\left(\gamma\frac{g_\mathrm{B}}{g_\mathrm{BF}}-\frac{\alpha_\mathrm{B}}{\alpha_\mathrm{F}}\right)+\frac{\alpha_\mathrm{B}}{\alpha_\mathrm{F}}\mu_\mathrm{F}-\mu_\mathrm{B}>0&
\end{flalign*}
}{
\hspace{-1pt}
\begin{flalign*}
&\chi\left(1-\gamma\right)<\left(\mu_\mathrm{F}-\frac{g_\mathrm{BF}}{g_\mathrm{B}}\mu_\mathrm{B}\right) &\\
&\frac{g_\mathrm{BF}}{g_\mathrm{B}}\alpha_\mathrm{B}-\alpha_\mathrm{F}<0,\quad\mu_\mathrm{F}-\frac{\alpha_\mathrm{F}}{\alpha_\mathrm{B}}\mu_\mathrm{B}>0 &\\
&\chi\left(\gamma\frac{g_\mathrm{B}}{g_\mathrm{BF}}-\frac{\alpha_\mathrm{B}}{\alpha_\mathrm{F}}\right)+\frac{\alpha_\mathrm{B}}{\alpha_\mathrm{F}}\mu_\mathrm{F}-\mu_\mathrm{B}>0&\\
& \mu_\mathrm{F}-\frac{\alpha_\mathrm{F}}{\alpha_\mathrm{B}}\mu_\mathrm{B} <\beta(1-\gamma)&\\
& g_\mathrm{B}\mu_\mathrm{F}-g_\mathrm{BF}\mu_\mathrm{B}+\beta\left(g_\mathrm{BF}\gamma\frac{\alpha_\mathrm{B}}{\alpha_\mathrm{F}}-g_\mathrm{B}\right)>0 &\\
\end{flalign*}
}
\vspace{-0.5cm}
\end{shaded}
\normalsize

\subsubsection{Fermionic scenario Fig.~\ref{FermionicScenarios}.c and Fig.~\ref{FermionicScenarios}.d}
\hspace{1cm}\\
The next step is to consider the fermionic scenarios Fig.~\ref{FermionicScenarios}.c and \ref{FermionicScenarios}.d but clearly only the latter leads to a physical configuration. Hence,
$\frac{g_\mathrm{BF}}{g_\mathrm{B}}\alpha_\mathrm{B}-\alpha_\mathrm{F}<0$ and a ceasing as well as tangent point exist.  When we require the ceasing point to be greater and the tangent point to be smaller than zero by means of
\Eqref{conditionForCeasing2} and \Eqref{conditionConnection3}, we find a second set of parameters for scenario 2.1. If, in contrast, the ceasing and tangent point are both smaller than zero, we have to require
a stationary point and it has to be above zero. Using \Eqref{CondStationary3} and \Eqref{CondStationary4}, this leads to a second set of parameters for scenario 2.2.

\subsubsection{Fermionic scenario Fig.~\ref{FermionicScenarios}.e and Fig.~\ref{FermionicScenarios}.f}
\hspace{1cm}\\
It is left to investigate the case of Fig.~\ref{FermionicScenarios}.e and Fig.~\ref{FermionicScenarios}.f but clearly only the former can lead to a physical situation as the solution in 
Fig.~\ref{FermionicScenarios}.f  is unstable everywhere. Thus, $\frac{g_\mathrm{BF}}{g_\mathrm{B}}\alpha_\mathrm{B}-\alpha_\mathrm{F}>0$. 
As the positive slope of the stable
fermionic solution branch prohibits a bosonic stationary point, the ceasing point has to be larger and the tangent point smaller than zero, ensured by \Eqref{conditionForCeasing2} and \Eqref{conditionConnection3}.
Therefore,  we arrive at scenario 3.

\footnotesize
\begin{shaded}
\noindent
3. $g_\mathrm{BF}>0$, zero fermionic density at the origin
\begin{equation*}
\vspace{-0.15cm}
\mu_\mathrm{F}-\frac{g_\mathrm{BF}}{g_\mathrm{B}}\mu_\mathrm{B}<0
\end{equation*}
\begin{equation*}
\vspace{-0.15cm}
\mu_\mathrm{F}-\frac{\alpha_\mathrm{F}}{\alpha_\mathrm{B}}\mu_\mathrm{B}>0,\quad \frac{g_\mathrm{BF}}{g_\mathrm{B}}\alpha_\mathrm{B}-\alpha_\mathrm{F}>0
\end{equation*}
\begin{equation*}
\vspace{-0.15cm}
\chi \left(\gamma\frac{g_\mathrm{B}}{g_\mathrm{BF}}-\frac{\alpha_\mathrm{B}}{\alpha_\mathrm{F}}\right)+\frac{\alpha_\mathrm{B}}{\alpha_\mathrm{F}}\mu_\mathrm{F}-\mu_\mathrm{B}<0
\end{equation*}
	
\vspace{-0.2cm}
\begin{figure}[H]
\begin{center}
\hspace{-0.7cm}
\includegraphics{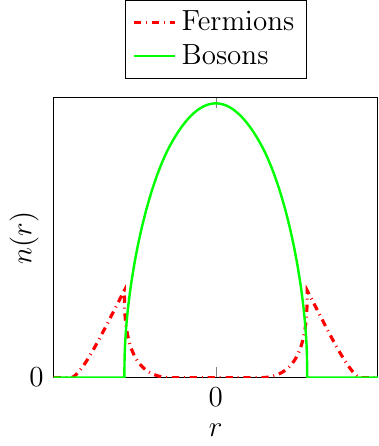}
\end{center}
\end{figure}
%%%%%%%%%%%%%%%%%%%%%%%%%%%%%%%%%%%%%%%%%%%%%%%%%%%%%%%%%%%%%%%%%%%%%%%%%%%%%%%%%%%%%%%%%%%%%%%%%%%%%%%%%%%%%%%%%%%%%%%%%%%
\vspace{-1cm}
\end{shaded}
\normalsize

\subsection{Classification for attractive interaction $g_\mathrm{BF}<0$}
\label{Classification for attractive interaction}
We now turn to the situation where $g_\mathrm{BF}<0$. While the geometric arguments we used to obtain the fermionic solutions in \Figref{FermionicScenarios} are still valid, the discussion in
\ref{Classification of bosonic solutions at the origin} is not due to the change of sign 
of $g_\mathrm{BF}$ in \Eqref{Decoupled22Origin}. As it will turn out, there are only two distinct scenarios here, which 
we will be  able to obtain by the results of \ref{Classification of bosonic solutions Vg0} that are valid for  both signs of $g_\mathrm{BF}$. It is important 
to note that according to \Eqref{WelcheSolutionIsStable} now the lower fermionic and the lower bosonic solution correspond to the stable density, whereas the upper bosonic solution 
is now unstable. From the conditions in \Figref{FermionicScenarios} we recognize that $g_\mathrm{BF}<0$ is only compatible with Fig.~\ref{FermionicScenarios}.b, d and f but obviously
Fig.~\ref{FermionicScenarios}.f does not lead to a physical configuration.
In these cases $\frac{g_\mathrm{BF}}{g_\mathrm{B}}\alpha_\mathrm{B}-\alpha_\mathrm{F}<0$ is always true. 
Recall from the discussion in \ref{Classification of bosonic solutions Vg0} that if $g_\mathrm{BF}<0$, there is no stationary point.\\
We will first investigate the bosonic solution corresponding to Fig.~\ref{FermionicScenarios}.d. This solution shows a tangent point and a ceasing point. In order to result in a reasonable density,
 we have to require that the bosonic tangent point is smaller than zero. When we recall that the slope of the bosonic solution branch at the ceasing point is negative due to \Eqref{SlopeCeasing} and that there 
is no stationary point, it follows that $n_\mathrm{B}<0$ 
everywhere and we therefore conclude that Fig.~\ref{FermionicScenarios}.d for $g_\mathrm{BF}<0$ does not lead to a physical density distribution. 

\footnotesize
\begin{shaded}
\noindent
4. $g_\mathrm{BF}<0$
\begin{equation}
\label{condition collapse}
\chi\left(1-\gamma\right)>\left(\mu_\mathrm{F}-\frac{g_\mathrm{BF}}{g_\mathrm{B}}\mu_\mathrm{B}\right)
\end{equation}

\flMinipage{0.59}{0.41}{
\hspace{2cm}
4.1 Fermionic density reaches zero first   
\vspace{-0.2cm}
\begin{flalign*}
\hspace{2.13cm}
&\mu_\mathrm{F}-\frac{\alpha_\mathrm{F}}{\alpha_\mathrm{B}}\mu_\mathrm{B}<0&
\end{flalign*}
}{4.2 Bosonic density reaches zero first
\vspace{-0.2cm}
\begin{flalign*}
&\mu_\mathrm{F}-\frac{\alpha_\mathrm{F}}{\alpha_\mathrm{B}}\mu_\mathrm{B}>0&
\end{flalign*}
}

\vspace{-0.5cm}
\begin{figure}[H]
	\hspace{1.92cm}
\includegraphics{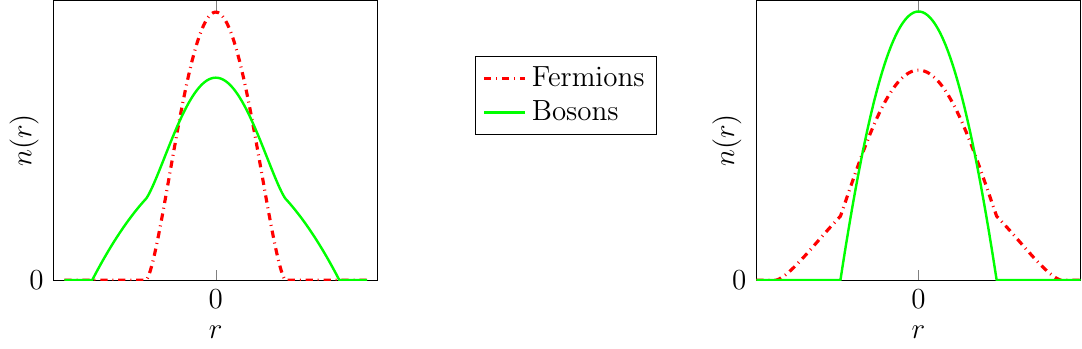}
\end{figure}
%%%%%%%%%%%%%%%%%%%%%%%%%%%%%%%%%%%%%%%%%%%%%%%%%%%%%%%%%%%%%%%%%%%%%%%%%%%%%%%%%%%%%%%%%%%%%%%%%%%%%%%%%%%%%%%%%
\vspace{-0.5cm}
\end{shaded}
\normalsize

\noindent
If the conditions of Fig.~\ref{FermionicScenarios}.b are fulfilled we observe a ceasing point. We first show that $n_\mathrm{B}>0$ at the origin. To this end, we cast \Eqref{EqsForGroundstate1}
in the form
\begin{equation}
n_\mathrm{B}=\frac{1}{g_\mathrm{B}}\left(\mu_\mathrm{B}+\vert g_\mathrm{BF}\vert n_\mathrm{F}\right)>0\;.  
\end{equation}
Depending on whether we choose the ceasing point above or below zero, we find two distinct configurations, namely scenario 4.1 and 4.2\\
We show in \ref{AppendixNoPhaseSeparation} that neither full nor partial phase separation can occur in the case of attractive interaction. Thus, if the parameters do not allow the continuous
configurations 4.1 or 4.2, there is no stable configuration possible and the system collapses, which happens exactly if \Eqref{condition collapse} is violated. This condition has already been obtained by 
Refs. \cite{55, 69} in order to determine the unstable regime in parameter space.\\
It is important to note that all the configurations described in this section are minima of the grand-canonical energy. Stability in the mixed region is guaranteed by selecting the
stable solution branch in \Secref{Stability}. We show in \ref{Stability in pure regions} that the configurations are also stable with respect to density variations in the pure regions.

\newpage
\section{Conclusion}
We have presented a complete classification of all continuous ground-sate density configurations of a boson-fermion mixture at zero temperature. 
 The results of this paper are valid for spin-polarized fermions or for a superfluid mixture at unitarity. We have assumed the bosons to interact among themselves 
and with the  fermions via s-wave scattering. In contrast, the interaction between the fermions can be neglected at extremely low temperatures. The system is
 trapped by a potential, which can be quite general, in particular it is not
 necessarily harmonic: It only has to be continuous and monotonous along every direction. We have started from the mean-field  grand-canonical energy functional,
 where the local-density approximation for fermions has been employed. In addition, we have neglected the bosonic kinetic-energy term in the spirit of the
 Thomas-Fermi approximation. It has been shown that the binary form of the system allows three different regions,  which have to be treated separately: 
A mixed region, where the bosonic and fermionic densities overlap, and two pure regions where only one species is present.  We have referred to the possible 
density configurations as `fully phase separated', if no mixed region exists, here one expects in general discontinuities in the densities, `partially phase separated'
 if a mixed region exists but there is at least one 
discontinuity, and `continuous' if the densities are continuous. We have briefly discussed phase separation in the context a mixture with attractive inter-species interaction, where we
 showed that phase separated configurations are not stable. This explains the onset of collapse in these systems when the choice of parameters 
do not allow continuous configurations.\\ In this paper we have investigated in detail continuous density profiles, which led to numerous scenarios. 
By taking the first variation of the grand-canonical energy functional, we have obtained coupled equations for the bosonic and fermionic density, respectively.
 The equations could be decoupled but are still nonlinear in the mixed region, giving rise to in general several solution branches. Remarkably, we have shown that 
the stability conditions for the coexisting region, resultant from the second variation of the energy functional, can be used to select, if existent, the stable
 solution branch, which is then unique. In the following course of constructing the density shapes we have pursued a geometric approach, that is, by trying to
 understand the behavior of the equations, conditions could be derived to first determine six different classes of fermionic density shapes. By additionally including
 analytic results about, for instance, the existence of maxima in the bosonic density or which species encloses the other, we were also able to establish the
 shape of the bosonic densities in the coexisting region. These could then be complemented by the solutions in the pure regions to finally obtain in total ten different configurations. Finally,
 we have proved that every continuous configuration constructed in this manner is also stable with respect to density variations in the  pure regions.\\ To our
 knowledge scenario 2.2 has not yet been described in the literature. This configurations consists of a shell of bosons completely enclosed by the fermionic 
cloud. It is the only density shape with two distinct pure regions of the same species. This feature is a manifestation of the nonlinearity of the equation
 as discussed in \ref{Stability in pure regions}. Our results can also be applied to a binary mixture of bosons by taking the limit $\gamma\rightarrow 1$
 in all the conditions. It is worthwhile noting that most of the configurations we find for the boson-fermion mixture are also possible in a binary mixture of bosons. There are two exceptions namely 
scenario 1.1.2 and 2.2 which cannot occur in this  case. \\ In conclusion, we have developed a rigorous and comprehensive classification scheme for the 
ground-state density profiles of a boson-fermion mixture with spin-polarized fermions and we have analytically proved stability for all the configurations that we describe.

\section*{Acknowledgment}
We thank M. Efremov, W. Zeller, S. Kleinert, J. Jenewein and A. Friedrich for many fruitful discussions. The QUANTUS project is supported by the German Space Agency DLR with
funds provided by the Federal Ministry for Economic Affairs and Energy
(BMWi) under grant number 50WM1556. 

\appendix

\section{Qualitative shape of the bosonic solutions for $g_\mathrm{BF}>0$}
\label{Qualitative shape of bosonic solutions}
As described in \Secref{FullCalssi}, the form of \Eqref{DecoupledGeneral2} does not permit to predict the qualitative shape of the bosonic solutions by the same simple arguments as
used in \Secref{Classification of fermionic solutions} for the fermionic classification. Complementing the discussion with the additional analytic conditions derived in this appendix,
 we will be able to determine the bosonic density shapes as well.
 First, in the case of two solutions at the origin their sign is determined in \ref{Classification of bosonic solutions at the origin}, followed by \ref{Classification of bosonic solutions Vg0} where we determine the values of
the bosonic density at ceasing, tangent, and stationary points.

\subsection{Sign of bosonic solutions at the origin ($V=0$)}\label{Classification of bosonic solutions at the origin}
\begin{figure}[H]
\begin{center}
\includegraphics{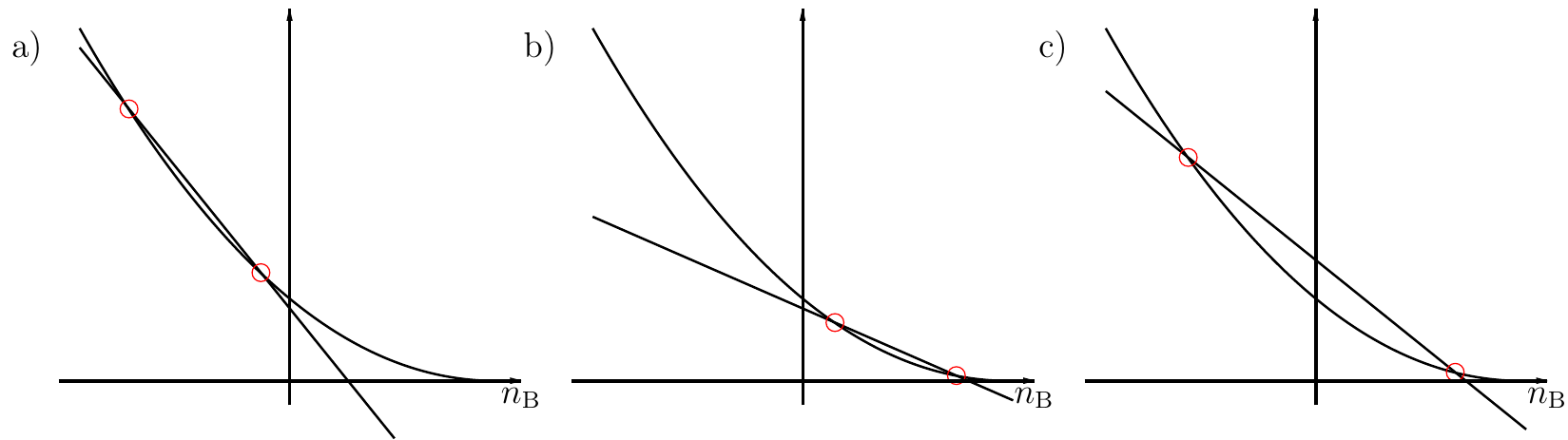}
\begin{alignat}{6}
\label{ConditionsThreeSolutions}
&\frac{\kappa^\frac{1}{\gamma}\mu_\mathrm{B}}{g_\mathrm{BF}}&&<\mu_\mathrm{F}^\frac{1}{\gamma}\qquad\qquad\qquad&&
\frac{\kappa^\frac{1}{\gamma}\mu_\mathrm{B}}{g_\mathrm{BF}}&&<\mu_\mathrm{F}^\frac{1}{\gamma} \qquad\qquad\quad\quad\qquad&&
\frac{\kappa^\frac{1}{\gamma}\mu_\mathrm{B}}{g_\mathrm{BF}}>\mu_\mathrm{F}^\frac{1}{\gamma}\\\nonumber
&\frac{1}{\gamma}\mu_\mathrm{F}^{\frac{1}{\gamma}-1}&&<\frac{\kappa^\frac{1}{\gamma}g_\mathrm{B}}{g_\mathrm{BF}^2} \qquad\qquad\qquad&&\frac{1}{\gamma}\mu_\mathrm{F}^{\frac{1}{\gamma}-1}&&>\frac{\kappa^\frac{1}{\gamma}g_\mathrm{B}}{g_\mathrm{BF}^2} &&   \qquad\qquad\qquad&& 
\end{alignat}
\caption{Geometric solution to \Eqref{DecoupledGeneral2}. Depending on the conditions \Eqref{ConditionsThreeSolutions}, 
two negative (Fig.~\ref{BosonenScenario1}.a), two positive (Fig.~\ref{BosonenScenario1}.b), or one positive and one negative bosonic solutions (Fig.~\ref{BosonenScenario1}.c) emerge at the origin.}
\label{BosonenScenario1}
\end{center}
\end{figure}

At the origin $V=0$ and \Eqref{DecoupledGeneral2} takes the simpler form
\begin{equation}
\label{Decoupled22Origin}
\frac{\kappa^\frac{1}{\gamma}\mu_\mathrm{B}}{g_\mathrm{BF}}-\frac{\kappa^\frac{1}{\gamma}g_\mathrm{B}}{g_\mathrm{BF}}n_\mathrm{B}=\left(\mu_\mathrm{F}-g_\mathrm{BF}n_\mathrm{B}\right)^\frac{1}{\gamma}\;.   
\end{equation}
This equation exhibits zero, one, or two real solutions. The case studied in this section is two solutions for which 
\Eqref{Solutionsattheorigin2} with $V=0$ provides the condition
\begin{equation}
\label{TwoRealSolutions2}
\chi\left(1-\gamma\right)>\left( \mu_\mathrm{F}-\frac{g_\mathrm{BF}}{g_\mathrm{B}}\mu_\mathrm{B}\right)>0\;.
\end{equation}
The left-hand side of \Eqref{Decoupled22Origin} is interpreted as a straight line with negative slope 
(for $g_\mathrm{BF}>0$), the right-hand side as a convex function. The solutions for $n_\mathrm{B}$ at the origin are again given by the intersection points of these two functions. 
Depending on the sign of the solutions, there can be three subcases, i.e.~two negative, two positive, or one positive and one negative solution.
These situations are illustrated in Fig.~\ref{BosonenScenario1}.a-\ref{BosonenScenario1}.c. 
In the following we derive conditions to distinguish these three cases by comparing the ordinate intersection of the two functions, given by $\frac{\kappa^\frac{1}{\gamma}\mu_\mathrm{B}}{g_\mathrm{BF}}$ 
for the straight line and
by $\mu_\mathrm{F}^\frac{1}{\gamma}$ for the convex function as can be seen from \Eqref{Decoupled22Origin} by setting $n_\mathrm{B}=0$. Thus,
if  $\frac{\kappa^\frac{1}{\gamma}\mu_\mathrm{B}}{g_\mathrm{BF}}<\mu_\mathrm{F}^\frac{1}{\gamma}$ the straight line crosses the ordinate below the convex function e.g.~in Fig.~\ref{BosonenScenario1}.a and \ref{BosonenScenario1}.b, while, in contrast, for $\frac{\kappa^\frac{1}{\gamma}\mu_\mathrm{B}}{g_\mathrm{BF}}>\mu_\mathrm{F}^\frac{1}{\gamma}$
 we observe the behavior depicted in Fig.~\ref{BosonenScenario1}.c.\\
In the case of $\frac{\kappa^\frac{1}{\gamma}\mu_\mathrm{B}}{g_\mathrm{BF}}<\mu_\mathrm{F}^\frac{1}{\gamma}$, there are either two negative (Fig.~\ref{BosonenScenario1}.a) or
 two positive solutions (Fig.~\ref{BosonenScenario1}.b).
In order to distinguish between these two situations, we compare the slope of the straight line with the slope of the convex function at $n_\mathrm{B}=0$ and find
two positive solutions if the slope of the straight line is greater than the first derivative of the convex function, which means
\begin{equation}
\frac{1}{\gamma}\mu_\mathrm{F}^{\frac{1}{\gamma}-1}>\frac{\kappa^\frac{1}{\gamma}g_\mathrm{B}}{g_\mathrm{BF}^2}\;.
\end{equation}

\noindent
Similarly, we find two negative solutions if the slope of the straight line is smaller than the first derivative of the convex function, i.e.~
\begin{equation}
\frac{1}{\gamma}\mu_\mathrm{F}^{\frac{1}{\gamma}-1}<\frac{\kappa^\frac{1}{\gamma}g_\mathrm{B}}{g_\mathrm{BF}^2}\;.
\end{equation}
Below \Figref{BosonenScenario1} we have summarized the additional conditions to specify the sign of the bosonic solutions at the origin in the case of two real solutions.

\subsection{Values of $n_\mathrm{B}$ at ceasing, tangent, and stationary points}\label{Classification of bosonic solutions Vg0}
We complete this appendix by deriving conditions for the existence and the values of ceasing, tangent, and stationary points in the bosonic solutions.
\vspace{0.5cm}\\
\noindent
\textit{Ceasing points:}
\hspace{1cm}\\
When both functions e.g.~in Fig.~\ref{BosonenScenario1}.a move to the left as $V$ is varied, but the straight line is `slower', there exists a $V^{(0)}$ where the upper solution branch ceases and there is no solution for $V>V^{(0)}$ anymore.
This happens exactly when the right-hand side and the left-hand side of \Eqref{DecoupledGeneral2} are simultaneously zero at the same point 
$n_\mathrm{B}^{(0)}$, giving rise to the system of  equations
\begin{alignat}{1}
\label{conditionForCeasing1.1}
&\alpha_\mathrm{B} V^{(0)}\;+\; g_\mathrm{B}\; n_\mathrm{B}^{(0)}-\mu_\mathrm{B}=0\\
\label{conditionForCeasing1.2}
&\alpha_\mathrm{F} V^{(0)}\;+\; g_\mathrm{BF} n_\mathrm{B}^{(0)}-\mu_\mathrm{F}=0
\intertext{with the solution}
\label{conditionForCeasing2}
&n_\mathrm{B}^{(0)}=\frac{\alpha_\mathrm{B}}{g_\mathrm{B}}\;\frac{\mu_\mathrm{F}-\frac{\alpha_\mathrm{F}}{\alpha_\mathrm{B}}\mu_\mathrm{B}}{\frac{g_\mathrm{BF}}{g_\mathrm{B}}\alpha_\mathrm{B}-\alpha_\mathrm{F}}\,.
\intertext{When we take the derivative of  \Eqref{DecoupledGeneral2} with respect to $V$ at the ceasing point, we find}
&\label{SlopeCeasing}
 \frac{\partial n_\mathrm{B}^{(0)}}{\partial V}=-\frac{\alpha_\mathrm{B}}{g_\mathrm{B}}\,.
\end{alignat}
Therefore, the bosonic density decreases at a ceasing point in the direction of increasing $V$.
Moreover, it is worthwhile noting that  if a ceasing point exists, it is in the upper bosonic solution as can easily be seen from \Figref{BosonenScenario1}.\\

\noindent
\textit{Tangent points:}
\hspace{1cm}\\
If the straight line moves 
`faster', the two intersection points approach each other until they combine. At this point $V^{(t)}$ the straight line is tangent to the convex function. Here, the 
two solution branches connect and there is no real solution for $V>V^{(t)}$. Thus, in addition to satisfying \Eqref{DecoupledGeneral2} we require the derivative 
with respect to $n_\mathrm{B}$ of its left-hand side and the derivative of its right-hand side to be equal at constant $V$. After rearranging terms, this reads
\begin{alignat}{1}
\label{conditionConnection1}
&\chi=\mu_\mathrm{F}-V^{(t)}_\mathrm{F}-g_\mathrm{BF}n_\mathrm{B}^{(t)}\;,
\intertext{where we have recalled \Eqref{Definechi}. Again, \Eqref{DecoupledGeneral2} and \Eqref{conditionConnection1} constitute  a system of equations promoting the solution }
\label{conditionConnection3}
& n_\mathrm{B}^{(t)}=\frac{\alpha_\mathrm{F}}{g_\mathrm{B}}\;\frac{\chi\left(\gamma\frac{g_\mathrm{B}}{g_\mathrm{BF}}-\frac{\alpha_\mathrm{B}}{\alpha_\mathrm{F}}\right)+\frac{\alpha_\mathrm{B}}{\alpha_\mathrm{F}}\mu_\mathrm{F}-\mu_\mathrm{B}}{    \frac{g_\mathrm{BF}}{g_\mathrm{B}}\alpha_\mathrm{B}-\alpha_\mathrm{F}}\;.
\end{alignat}

\noindent
\textit{Stationary points:}
\hspace{1cm}\\
In order to completely capture the behavior of the bosonic solutions, we have to investigate if $n_\mathrm{B}$ exhibits stationary points, that is, if there exists a $V^{(s)}>0$ with $\partial n_\mathrm{B}^{(s)}/\partial V=0$.
When we take the derivative with respect to $V$ on both sides of \Eqref{DecoupledGeneral2} and set $\partial n_\mathrm{B}^{(s)}/\partial V=0$, we obtain
\begin{equation}
\label{CondStationary1}
\frac{\alpha_\mathrm{B}}{\alpha_\mathrm{F}}\gamma\frac{\kappa^\frac{1}{\gamma}}{g_\mathrm{BF}}=\left(\mu_\mathrm{F}-V_\mathrm{F}^{(s)}-g_\mathrm{BF} n_\mathrm{B}^{(s)}\right)^{\frac{1}{\gamma}-1}\;.
\end{equation}
Clearly, there is no solution to this equation if $g_\mathrm{BF}<0$, thus there will be no stationary point in the case of attractive inter-species interaction. For repulsive interaction the system of \Eqref{DecoupledGeneral2} and \Eqref{CondStationary1} admits the solution
\begin{equation}
\label{CondStationary3}
n_\mathrm{B}^{(s)}=\frac{\alpha_\mathrm{B}}{g_\mathrm{B}\left(\frac{g_\mathrm{BF}}{g_\mathrm{B}}\alpha_\mathrm{B}-\alpha_\mathrm{F}\right)}  \left[\mu_\mathrm{F}-\frac{\alpha_\mathrm{F}}{\alpha_\mathrm{B}}\mu_\mathrm{B} -\beta\left(1-\gamma\right)\right]\;.
\end{equation}
Here we have defined the parameter
\begin{equation}
\label{DefBeta}
\beta:=\kappa\left(\gamma\frac{\alpha_\mathrm{B}}{\alpha_\mathrm{F}}\frac{\kappa}{g_\mathrm{BF}}\right)^\frac{\gamma}{1-\gamma}\;.
\end{equation}
Thus, a stationary point additionally to the one at the origin exist if
\begin{equation}
\label{CondStationary4}
-V^{(s)}=\frac{1}{g_\mathrm{B}\left(\frac{g_\mathrm{BF}}{g_\mathrm{B}}\alpha_\mathrm{B}-\alpha_\mathrm{F}\right)}  \left[g_\mathrm{B}\mu_\mathrm{F} -g_\mathrm{BF}\mu_\mathrm{B}+\beta\left(g_\mathrm{BF}\gamma  \frac{\alpha_\mathrm{B}}{\alpha_\mathrm{F}}-g_\mathrm{B}\right)\right]<0\,.
\end{equation}
When we take the derivative with respect to $V$ of \Eqref{EqsForGroundstate1} and set $\partial n_\mathrm{B}^{(s)}/\partial V=0$ at the stationary point, we arrive at
\begin{equation}
\label{StationaryPointSlopeCondition}
0=\alpha_\mathrm{B}+g_\mathrm{BF}\frac{\partial n_\mathrm{F}^{(s)}}{\partial V}
\end{equation}
Obviously, for positive $g_\mathrm{BF}$ this can only be true when the fermionic density decreases as $V$ is increased. This argument therefore
determines which solution branch becomes stationary.\\

\section{Stability in $\Omega_\mathrm{B}$ and $\Omega_\mathrm{F}$}
\label{Stability in pure regions}
\begin{figure}[h]
	\begin{center}
\includegraphics{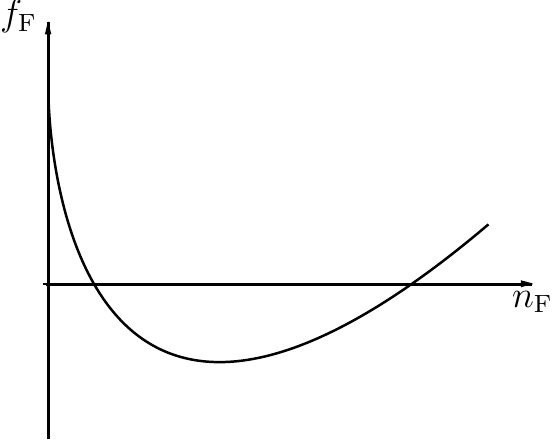}
		\caption{Qualitative shape of $f_\mathrm{F}$ in the case of two zeros. }
		\label{StabilityPureRegions}
	\end{center}
\end{figure}

In this appendix we show that every scenario with a mixed region and continuous densities is stable with respect to the inequalities \Eqref{AdditionalCond1} and (\ref{AdditionalCond2}) 
in the pure region. We then conclude that every scenario derived in \Secref{FullCalssi} is stable.\\
We start by defining
\begin{alignat}{3}
\label{Definitionfb}
&f_\mathrm{B}  &&:=V_\mathrm{F}+g_\mathrm{BF}n_\mathrm{B}-\mu_\mathrm{F}\quad\quad\text{and}\\
\label{Definitionff}
&f_\mathrm{F}  &&:=V_\mathrm{B}+g_\mathrm{BF}n_\mathrm{F}-\mu_\mathrm{B}\;,
\intertext{so that it remains to show that}
\label{ConditionZusammengefasst}
&f_i &&\geq0\;\;\text{in}\;\;\Omega_i\;\;\text{for}\;\;i=\mathrm{B,F}.
\intertext{Note that the functions $f_i$ are continuous as the potentials and the densities are continuous.
Comparing \Eqref{Definitionfb} and \Eqref{Definitionff} with \Eqref{EqsForGroundstate1} and \Eqref{EqsForGroundstate2} leads to}
\label{conditioninmixedregion1}
&f_\mathrm{B}  &&=-\kappa n_\mathrm{F}^\gamma\quad\quad\text{and}\\
\label{conditioninmixedregion2}
&f_\mathrm{F}  &&=-g_\mathrm{B}n_\mathrm{B}
\end{alignat}
for $\mathbf{r} \in \Omega_\mathrm{BF}$. Thus, $f_i$ is smaller than zero in the mixed region and $f_i=0$ at the border between 
$\Omega_\mathrm{BF}$ and $\Omega_i$ by definition of the pure region. Consequently, $f_i$ switches sign from negative to positive at the border between the mixed and the pure region and thus it is left to show that $f_i\neq 0$ everywhere else in $\Omega_i$.
Before we prove that this is indeed true, it is important to note that $n_i$ is a monotonous function over the union of all pure regions of species $i$ in the same sense of monotony as defined  
in \Secref{Minima of the grand-canonical energy}. This statement follows from \Eqref{EqsForGroundstate3} and \Eqref{EqsForGroundstate4} as the potentials are monotonous.
As a consequence, it is sufficient to consider $f_i$ a function of $n_i$ instead of $\mathbf{r}$.\\
By writing $V_\mathrm{B}=\alpha_\mathrm{B}V$ and $V_\mathrm{F}=\alpha_\mathrm{F}V$, we insert \Eqref{EqsForGroundstate3} into \Eqref{Definitionfb} and \Eqref{EqsForGroundstate4} 
into \Eqref{Definitionff} to eliminate the potentials. This leads to the expressions
\begin{alignat}{1}
\label{Fulldefinitionstability1}
f_\mathrm{B} & =-\left(\mu_\mathrm{F}-\frac{\alpha_\mathrm{F}}{\alpha_\mathrm{B}}\mu_\mathrm{B}\right)+\frac{g_\mathrm{B}}{\alpha_\mathrm{B}}\left(\frac{g_\mathrm{BF}}{g_\mathrm{B}}\alpha_\mathrm{B}-\alpha_\mathrm{F}\right)n_\mathrm{B}\\
\intertext{and}
\label{FullDefinitionStability2}
f_\mathrm{F} & =\frac{\alpha_\mathrm{B}}{\alpha_\mathrm{F}}\left(\mu_\mathrm{F}-\frac{\alpha_\mathrm{F}}{\alpha_\mathrm{B}}\mu_\mathrm{B}+\frac{\alpha_\mathrm{F}}{\alpha_\mathrm{B}}g_\mathrm{BF}n_\mathrm{F}-\kappa n_\mathrm{F}^\gamma\right)\;,
\end{alignat} 
valid in the pure bosonic and fermionic regions, respectively.\\ \Eqref{Fulldefinitionstability1} is a linear function of $n_\mathrm{B}$ and therefore admits one or no zero.
 If a border between $\Omega_\mathrm{BF}$
and $\Omega_\mathrm{B}$ exists, then $f_\mathrm{B}$ is negative in $\Omega_\mathrm{BF}$,  zero at the border, and consequently positive in $\Omega_\mathrm{B}$, hence
we have proved \Eqref{ConditionZusammengefasst} for $i=\mathrm{B}$.\\ The argument for $f_\mathrm{F}$, on the other hand, is slightly more involved.
 Nevertheless, when we realize 
that the right-hand side of \Eqref{FullDefinitionStability2}
is of the same form as \Eqref{DecoupledGeneral1}, provided that $g_\mathrm{BF}>0$, we are able to apply the considerations made in \Secref{Conditions for the number of solutions} 
according to which a function
of this form can have no, one or two zeros. In the case of one zero the pure 
fermionic region is stable 
according to the same argument as used for $f_\mathrm{B}$. In the case of two zeros, let us determine how $f_\mathrm{F}$ behaves for large $n_\mathrm{F}$. As $\gamma<1$,
\begin{equation}
f_\mathrm{F}\rightarrow \infty\quad\quad \text{for}\quad\quad n_\mathrm{F}\rightarrow\infty\;,
\end{equation}
that is, the function qualitatively looks like as depicted in \Figref{StabilityPureRegions}. Note that $f_\mathrm{F}(n_\mathrm{F}=0)>0$ since we consider the case of two zeros.
Given a border between $\Omega_\mathrm{BF}$ and $\Omega_\mathrm{F}$, it might correspond to the first or second zero of $f_\mathrm{F}$. But irrespectively of that, once a border from the mixed region
to the pure fermionic region is crossed, $f_\mathrm{F}$ stays larger than zero. It is also readily shown that $f_\mathrm{F}$ assumes no or one zero if $g_\mathrm{BF}<0$. Thus,  we have proved \Eqref{ConditionZusammengefasst} also for $i=\mathrm{F}$.
\noindent
In summary we have found that all density configurations which are constructed as described in the end of \Secref{Minima of the grand-canonical energy} are (meta-) stable. It is interesting to note that the results of this appendix only allow
at most one border between mixed and pure bosonic region but two borders between mixed and pure fermionic region in a given direction. This is exactly what we observe in 
\Secref{Classification for repulsive interaction} and \Secref{Classification for attractive interaction} where the only configuration with two borders between $\Omega_\mathrm{BF}$ and $\Omega_\mathrm{F}$
is scenario (2.2).

\section{Origin of the collapse in the case of $g_\mathrm{BF}<0$}
\label{AppendixNoPhaseSeparation}
In this appendix we show that $g_\mathrm{BF}<0$ precludes phase separation.\\ Clearly, in this case the components attract each other and the system therefore rather tries to
increase the overlap between the densities than tends to separate and demix. But what is clear from physical reasoning can also be shown rigorously.\\ We begin by noting that for $g_\mathrm{BF}<0$ \Eqref{Fulldefinitionstability1} and \Eqref{FullDefinitionStability2} are 
monotonically decreasing functions of the respective densities. Depending on the sign of $\mu_\mathrm{F}-\frac{\alpha_\mathrm{F}}{\alpha_\mathrm{B}}\mu_\mathrm{B}$, either $f_\mathrm{B}$ or $f_\mathrm{F}$  only assumes negative values
and therefore precludes a pure region for that species. Hence, scenarios with both a pure fermionic and a pure bosonic region do not exist and full phase separation cannot occur.\\
In the second step we prove that every partially phase separated scenario is unstable as well. To this end, let us assume a mixed region where at a certain point the density of one
component suddenly drops to zero, the density of the other species will then show a discontinuity. Exemplarily, we assume that the bosonic density drops to zero, the discontinuity will then emerge 
in the fermionic density. The proof in case of a bosonic
discontinuity goes along similar lines and will therefore not be shown here. Let us now denote all quantities at the discontinuity by the superscript $(d)$ when they correspond to the limit from the pure region and by $(m)$  for the limit from the mixed region. 
Evaluating  \Eqref{Definitionff} at both sides of the discontinuity and subtracting the results from each other yields
\begin{align}
\label{AusgangKeineDiscontinuity}
&f_\mathrm{F}^{(d)}-f_\mathrm{F}^{(m)}=g_\mathrm{BF}\left(n_\mathrm{F}^{(d)}-n_\mathrm{F}^{(m)}\right)\;.\\
\intertext{We also obtain} 
\label{AusgangKeineDiscontinuity2}
&n_\mathrm{F}^{\gamma\; (d)}  -n_\mathrm{F}^{\gamma\; (m)}=\frac{g_\mathrm{BF}}{\kappa}n_\mathrm{B}^{(m)}<0
\end{align}
from \Eqref{EqsForGroundstate2} and \Eqref{EqsForGroundstate4}. Note that from the equation above it follows that the density jumps to a greater value at the border from the pure to the mixed region.
With the help of the mean-value theorem we rewrite
\begin{equation}
\label{Meanvalue2}
n_\mathrm{F}^{\gamma\; (d)}  -n_\mathrm{F}^{\gamma\; (m)}  =\gamma\xi^{\gamma-1}\left(n_\mathrm{F}^{(d)}  -n_\mathrm{F}^{(m)} \right)<\gamma n_\mathrm{F}^{\gamma-1\;(m)}\left(n_\mathrm{F}^{(d)}  -n_\mathrm{F}^{(m)} \right)\;,
\end{equation}
where we have made use of the fact that $\xi\in(n_\mathrm{F}^{(d)},n_\mathrm{F}^{(m)} )$ and $n_\mathrm{F}^{(m)}>n_\mathrm{F}^{(d)}$. When we substitute \Eqref{conditioninmixedregion2},
into \Eqref{AusgangKeineDiscontinuity}, we obtain the relation
\begin{align}
\nonumber
 f_\mathrm{F}^{(d)}&=-g_\mathrm{B}n_\mathrm{B}^{(m)}+g_\mathrm{BF}\left(n_\mathrm{F}^{(d)}-n_\mathrm{F}^{(m)}\right)\\
\nonumber
                   &<-g_\mathrm{B}n_\mathrm{B}^{(m)}+\frac{g_\mathrm{BF}}{\gamma}n_\mathrm{F}^{1-\gamma\;(m)}\left(n_\mathrm{F}^{\gamma\;(d)}-n_\mathrm{F}^{\gamma\;(m)}\right)\\
\nonumber
                   &=g_\mathrm{B}n_\mathrm{B}^{(m)}n_\mathrm{F}^{1-\gamma\;(m)}\left(\frac{g_\mathrm{BF}^2}{\kappa\gamma g_\mathrm{B}}-n_\mathrm{F}^{\gamma-1\;(m)}\right)\\
		   &<0\;.
\end{align}
The first inequality follows from \Eqref{Meanvalue2} since $g_\mathrm{BF}<0$. We then use \Eqref{AusgangKeineDiscontinuity2} and 
the last inequality follows from \Eqref{Bedingung_nF}, that is, from the stability of the density in the mixed region.\\
In summary we have shown that in the case of attractive interaction there are neither stable fully phase-separated nor stable partially phase-separated configurations. When the conditions above scenario 4.1 and 4.2
are not satisfied, then there exists no other stable solution within our model. Indeed, the system collapses as has been experimentally observed by \cite{G}.

\newpage
\bibliographystyle{unsrt}

\end{document}